\documentclass[12pt, a4paper]{article}

 \usepackage[bottom]{footmisc}

\usepackage{graphicx}
\usepackage{amsthm}   
\usepackage{amsmath} 
\usepackage{amssymb}  
\usepackage{mathrsfs} 
\usepackage{stmaryrd} 
\usepackage{txfonts} 

\usepackage{hyperref}  
\usepackage[capitalize]{cleveref}
\usepackage{setspace}
\usepackage{enumerate}
\usepackage[all]{xy}

\usepackage{hyperref}
 \usepackage[font=itshape]{quoting}

\usepackage{mathtools}
\DeclarePairedDelimiter\anglebracket\langle\rangle

\newcommand{\mathsym}[1]{{}}
\newcommand{\unicode}[1]{{}}

\newcommand{\R}{{\mathbb{R}}}
\newcommand{\B}{{\mathbb{B}}}
\newcommand{\N}{{\mathbb{N}}}
\newcommand{\Z}{{\mathbb{Z}}}

\newcommand{\DD}{\mathcal{D}}

\newcommand{\SSS}{{\mathcal{S}}}

\newcommand{\TT}{{\mathcal{T}}}
\newcommand{\NN}{{\mathcal{N}}}
\newcommand{\WW}{{\mathcal{W}}}

\newcommand{\cs}[1]{\mathsf{#1}}   

\usepackage{color}  
\RequirePackage[dvipsnames,usenames]{xcolor}

\newcommand{\eq}[1]{\begin{align}#1\end{align}}

\newcommand{\be}{\begin{equation}}
\newcommand{\bel}[1]{\begin{equation}\label{#1}}
\newcommand{\qe}{\end{equation}}
\newcommand{\ee}{\end{equation}}
\newcommand{\eeq}{\end{equation}}
\newcommand{\ba}{\begin{eqnarray}}
\newcommand{\ea}{\end{eqnarray}}

\def\bal#1\eal{\begin{align}#1\end{align}}
\def\bann#1\eann{\begin{align*}#1\end{align*}}

\newtheorem{theorem}{Theorem}

\newtheorem{definition}{Definition}
\newtheorem{example}{Example}

\newtheorem{lemma}[theorem]{Lemma}

\date{}

\begin{document}

\title{Implications of computer science theory for the simulation hypothesis}

\author{David H. Wolpert \\
 Santa Fe Institute, 1399 Hyde Park Road, Santa Fe, NM, 87501\\
Complexity Science Hub, Vienna, Austria\\
International Center for Theoretical Physics, Trieste, Italy\\
Arizona State University, Tempe, AZ\\
\texttt{http://davidwolpert.weebly.com}
 }

\maketitle

\begin{abstract}
The simulation hypothesis has recently excited renewed interest, especially in the physics and philosophy communities.
However, the hypothesis specifically concerns {\textit{computers}} that simulate physical universes, which
means that to formally investigate it we need to couple computer science theory with physics. Here I 
couple those fields with the physical Church-Turing thesis. I then
exploit that coupling to investigate some of the computer science theory aspects of the simulation hypothesis.
In particular, I use Kleene's second recursion
theorem to prove that it is mathematically possible for {us} to be a simulation that is being run on a computer --- {by us}.
In such a self-simulation, there would be two identical instances of us; the question of which of those is ``really us'' is meaningless.
I also show how Rice's theorem provides some interesting impossibility results concerning simulation and self-simulation; 
briefly describe the philosophical implications of fully homomorphic encryption for (self-)simulation; and briefly investigate
the graphical structure of universes simulating universes simulating universes $\ldots$, among other issues. I end by describing some of the
possible avenues for future research.
\end{abstract}

\begin{quote}
\textit{He didn't know if he was Zhuang Zhou dreaming he was a butterfly, or a butterfly dreaming that he was Zhuang Zhou. }
\end{quote}

--- Zhuangzi, chapter 2 (Watson translation~\cite{watson2003zhuangzi})

$ $


\section{Introduction}

\subsection{Background}

Whatever the other dynamic properties of our physical universe, it is common to suppose that 
it can contain and run an arbitrary Turing machine.  In other words, it is common to suppose that the
dynamical system implemented by our universe can represent the dynamics of any Turing machine.
It is also broadly accepted that our universe's dynamics cannot
have properties that are not computable on a Turing machine. 

This second supposition raises the interesting hypothesis that we are a program being run on some 
Turing machine (or even more powerful computational machine). 
This possibility is known as the ``simulation hypothesis''. In one or form or another it
has existed for thousands of years, in many human cultures. 
Since the advent of powerful digital computers over the last two decades however, the simulation hypothesis has excited renewed
interest among physicists as well as philosophers of science.

In addition, whether or not one or the other the suppositions actually applies to our specific physical universe,
one can investigate the implications they would have if they do. In particular, one can investigate
what those two suppositions tell us about the relationship between any given dynamically evolving
universe and three phenomena:
\begin{enumerate}
\item The computational power that that system possesses;
\item The computational power that is needed to implement that dynamical system;
\item The implications of computer science (CS) theory for such a universe
\end{enumerate} 
The first two have previously been considered in the literature. However, the third
one is completely uninvestigated. It is the focus of this paper.

The modern iteration of the simulation hypothesis has many versions that vary in their precise details. 
Common to all of these versions is the idea that some portion of a physical
universe, including some conscious reasoning agents that exist in that universe, 
might in fact be part of a simulation that is being run in a physical computer of some different, super-sophisticated
reasoning agents, like some race of aliens~\cite{bostrom2003we,hamieh2021simulation,chalmers2022reality+}.{\footnote{A special 
case of this idea in which the ``super-sophisticated reasoning agents'' are our descendants, and so by simulating us they are simulating
their ancestors, was the focus of some of the recent work on the simulation hypothesis~\cite{bostrom2003we}. Here I impose
no such restrictions on the agents simulating us. Indeed, the physical universe in which the those reasoning agents exist might not
even be one that obeys the laws of physics of our (simulated) universe.}
Under the simulation hypothesis, those ``conscious reasoning agents'' would just be variables evolving 
in a computer program running in that physical computer designed by that
putative super-sophisticated race of aliens.

In particular, it might be that a portion of \textit{our} physical universe is being simulated this way, and
that in fact \textit{we} are just variables evolving in some computer in that physical universe.  
The vague implication is that if we are in fact a simulation, then what we {perceive} as reality does not
have any sort of ``objective truth'', since it is just a simulation. What {we} think of as ``reality''
would in fact be a chimera, simply a play being put on by the super-sophisticated race of aliens.
Crucially, that play could just as easily be put on by some other super-sophisticated race of aliens, 
in some other physical universe --- what we perceive as reality is just the play, not the physical universe in which the play
is being staged.

Much of philosophy of science, going back to Kant (at least),
would be rendered moot if we are simulations. In particular, this would be the case for many
of the flavors of ``*** realism'' (e.g., scientific realism), which suppose there ``exists'' some ``real'' physical truth
which is not reducible to a mathematical formalism, a physical truth that instead is somehow 
``concrete'', and non-mathematical in its very essence --- and so not just a computer algorithm.\footnote{Arguably, 
the simple fact that the simulation hypothesis is 
logically consistent, but if true would imply that there is no such ``concrete'', non-mathematical reality, establishes
that it is impossible to establish the logical necessity of any of these flavors of realism,
}

The central idea of the simulation hypothesis has been extended in an obvious manner, simply by noting that the
aliens that simulate our universe might themselves be a simulation
in the computer of some even more sophisticated species, and so on and so on,
in a sequence of ever-more sophisticated aliens. Similarly, going the other direction, in the
not too distant future we might produce our own simulation of a universe running in some future
computer that we will create, a simulation complete with variables that constitute ``conscious, reasoning agents''.
Indeed, we might produce such a simulation in which the reasoning agents can produce their own
simulated universe in turn, etc. So in the near future, there might be a sequence of species', each
one with a computer running a simulation that produces the species just below
it in the sequence, with us somewhere in the ``middle'' of that sequence.

The literature of the last few decades on the simulation hypothesis has focused
almost exclusively on whether {we, in our universe} {are} such a simulation.
This question is answered rather trivially if we adopt the view of ontic structural realism, especially if 
that view is formalized in terms of Tegmark’s level IV multiverse~\cite{tegmark1998theory,tegmark2008mathematical}: 
yes, in some universes we are a simulation, and no, in some other universes we are not. It is not an either / or.  'Nuf said.
(See also~\cite{carroll_wilczek_simulation_universe_2021,piccinini2011physical}.)

Even if we subscribe to the idea of a level IV multiverse though, there are
several research directions one can pursue. Some people have focused on ascribing probabilities to the specific hypothesis
that {we, in our universe}, are programs in such a simulation~\cite{bostrom2003we,chalmers2022reality+,kipping2020objective}.
In the language of the level IV multiverse, these researchers have asked what the relative probabilities 
are of the set of all universes in
which we are programs in a simulation and the set of all universes in which we are not. 

These investigations of relative probabilities all beg (many) questions,
about what such probabilities might actually mean, formally.
(Just as there are such questions concerning the idea of ascribing probabilities to the universes in a level IV multiverse
in general.) How does one ascribe a measure, obeying the Kolmogorov axioms based on some associated sigma algebra, to
a collection of events each of which is a universe? Note that it is not even clear that this collection
would be a set, rather than a proper class of some sort. Directly reflecting such problems, one could not experimentally assess 
any proposed value of such a probability, e.g., 
with a proper scoring rule.\footnote{One might try to use a Bayesian ``degree of belief'' interpretation of probability to
circumvent this issue. However, without any decision that would be made based on the probability assigned,
and associated loss function, one could not apply Bayesian decision theory. So it is hard to see how
``degree of belief'' is meaningful in this case.}
See~\cite{weatherson2003you} for some related arguments.

In particular, many papers considering such probabilities assume, implicitly or otherwise, that there is some
way to assign a uniform probability to all universes. It is proven below in \cref{sec:PCT} that this is mathematically
impossible in the kind of universes considered in this paper (which, one might argue, is the kind of universes considered
in these other papers). One might imagine that rather than a uniform probability distribution, one could assign some sort
of Cantor (``fair coin'') measure to the set of possible universes, as is done for example in algorithmic information theory. However,
it is proven in \cref{sec:simulation_graph} that it is impossible to assign a Cantor measure to 
a set of universes obeying a natural set of restrictions.

In contrast to this work on the probability that we are living in a simulation, some physicists have
focused on whether there might be ways of experimentally determining whether our
universe is a program in a simulation~\cite{campbell2017testing,beane2014constraints,barrow2007living}. 
Work in this vein can only provide partial
answers to the question of whether we are a simulation, at best.  For 
example, some of this work first make the assumption that the simulating computer is digital,
and so cannot fully faithfully represent real numbers. (And also assumes that the state space of our
universe is uncountably infinite, rather than being countable.) Other work in this vein investigates possible empirically
observable effects of ``bugs'' in the code being run by the simulating computer.

Yet others have considered issues concerning the simulation hypothesis that might charitably
be characterized as ``philosophical'' in nature~\cite{hanson2001live,conitzer2019puzzle,dainton2012singularities,lim2022we},
at least in that they do not involve mathematical analysis of any sort.

Note though that the simulation hypothesis explicitly asks whether one universe (e.g., ours)
might be a \textit{computer simulation} being run in some (perhaps different) universe. 
Strangely, despite its being formulated this way, there is almost no previous work investigating the simulation hypothesis
using the tools of computer science (CS) theory, or related fields like logic theory.

One earlier set of semi-formal results along these lines was presented in~\cite{wolpert2024know}. Some of those 
semi-formal results define ``simulation''
in terms of the relationship between languages at different levels of the time, polynomial, or exponential
hierarchies of CS theory, and some of them define ``simulation'' in terms of the relationship between languages at different levels
the arithmetic or analytic hierarchies of logic, or similar constructions. (See~\cite{cooper2017computability,arora2009computational,hajek1979arithmetical} for background on such hierarchies.) 
Other results in that paper define ``simulation'' in
terms of the relationship between computational machines with  different Turing degrees~\cite{soare2016Turing,shore2016Turing}.
Yet others consider the application of G{\"o}del's  second incompleteness theorem, defining simulation in terms
of languages (in the logic theory sense) with nested sets of axioms.

None of these earlier results in the literature consider what it means for a computational machine to simulate a \textit{physical universe}, per se.
There is no concern for ``coupling'' the mathematics of CS theory to the laws of physics of our universe.
One natural way to address this lacuna is to use a particular formalization of the physical Church-Turing thesis (PCT), 
and a closely related thesis which I call the reverse PCT (RPCT). That is the 
approach I adopt in this paper.\footnote{The reader unfamiliar with TMs and associated concepts like Universal Turing machines
(UTMs), instantaneous descriptions (ID) of the state of a TM and its tapes, prefix-free encodings, etc.,
should consult \cref{app:TMs} for appropriate background. } Specifically, in this paper I will say that the PCT
applies to a particular (portion of a) universe if the dynamics of that (portion of a) universe can be implemented
on a Turing machine (TM). The RPCT instead says that any desired TM can be implemented by the dynamics
of the universe in question, by appropriate choice of the initial conditions of that universe.

\subsection{General comments}

It is important to emphasize that in this paper I do \textit{not} assume that the PCT applies to our actual physical
universe. I do not even restrict attention to those universes that obey the laws of physics of our actual universe, as we currently
understand those laws. The focus is instead more general, considering what CS theory has to say about any 
universe containing a computer that runs a simulation of a universe. In particular,
in this paper I am only concerned with establishing the \textit{logical}
possibility of a physical system $V$ that can (contain a computer that can) simulate some universe $V'$.
I do not investigate the possibility of such a system under the laws of physics as currently understood
 --- never mind the even more narrowly defined question of whether there \textit{is} such a physical system in our
universe, evolving in a manner consistent with the laws of physics as currently understood.


It is also important to emphasize that although I will frequently refer to the computer in a universe as a (U)TM, 
I do not mean that it is a physical system consisting of a set of infinite tapes with associated heads, etc. Rather I just mean that
it has the properties of a (U)TM, i.e., that it is computationally universal. I then choose to discuss this
computational system as though it were implemented as a (U)TM. So for example, the universe could be
a laptop with a memory that can be extended dynamically an arbitrary finite amount
an arbitrary number of times as it runs. (See \cref{sec:example_V_our_universe} below
for a detailed example of how a subset of the specific universe occupied by us humans fits into this framework.)

Furthermore, there is no semantics in this paper, only syntax (in the terminology of the foundations of mathematics).
There are no structure functions, models, etc. Concretely, this means that there is no distinction between universes
that are ``real'' and those that are ``only simulations''. This reflects the viewpoint of the simulation hypothesis itself.
Formally, in this paper this lack of explicitly distinguishing real from non-real universes is possible due to
my exploiting the physical Church-Turing thesis.

\subsection{Contributions}

I begin in \cref{sec:preliminaries} by presenting the mathematical framework I will use in this paper.
I then use that to framework to formally define ``simulate'' in a way that can apply to any pair of dynamical systems,
with no restrictions to the laws of our particular universe. 
Next, I formally define the PCT in terms of the mathematical framework, as well as the RPCT.
These definitions provide the first formalization of the simulation hypothesis, as well as the the first fully generalization
formalization of the PCT, applicable to arbitrary universes, not just ours. These definitions also provide the 
first fully general distinction of the PCT from the RPCT.
To help the reader ground the discussion, I also sketch a way to relate the ``computers'' considered in these definitions to physical
subsystems of our actual universe. 

In the next section I explicitly 
prove that if a universe $V'$ obeys the PCT, and a universe $V$ obeys the RPCT, then $V$ can simulate
$V'$. I end that section by describing several arguments based on this result that one might suppose disprove
the possibility of self-simulation, i.e., which prove that we could not be simulations in a computer that we ourselves run.

In \cref{sec:self-sim-lemma} I use Kleene's second recursion theorem to address these
arguments.\footnote{This theorem
is just called ``the recursion theorem'' in CS theory; see~\cref{app:recursion_theorem-rice_theorem} for a summary.}
I use that theorem to prove that in fact we \textit{could} be simulations in a computer that {we ourselves run}.
Specifically, I show that if a universe $V$ obeys both the PCT and the RPCT,
then it can simulate itself, according to the formal definition of ``simulate'' provided in \cref{sec:framework}.
I call this the \textit{self-simulation lemma}. I then describe several important formal features of
the self-simulation lemma, and present an example of how self-simulation might arise with advanced
versions of our current laptops. I end this section by describing how self-simulation is a far deeper
connection between an entity and itself than arises in all the earlier versions of self-reference considered
in the mathematics literature.

In the following section I present several mathematical properties of the number of iterations taken
to simulate one's one dynamics a given time $\Delta t$ into the future. Some of these
involve requirements that the time taken to simulate $\Delta t$ into one's own
future does not decrease with $\Delta t$, for any specific pair of values of $\Delta t$. Other properties
involve the time-complexity of self-simulation, i.e.,
how much longer than a time $\Delta t$ it takes to simulate a universe's evolution up to a time $\Delta t$
in the future. 

The next section, \cref{sec:philo_implications} starts with a 
a discussion of some of the peculiar philosophical implications of the simulation lemma, and
especially of the self-simulation lemma, for notions of identity. In particular, that
section contains a discussion of the fact that 
self-simulation does not just mean that you create some doppelganger of yourself, a clone
of yourself, which has autonomy and starts to evolve differently from you once it
has been created. Self-simulation does not mean something akin to your stepping into a variant of the Star Trek
transporter which creates a copy of you at some other location while the original you still exists. 
Rather than such cloning of yourself, self-simulation
means that you run a program on a computer which implements the exact same dynamics 
as your \textit{entire} universe, the universe that contains both you and your computer. So in particular, that universe being simulated in
a program running in your computer $N$ contains an instance of you who, in this simulation, is running a 
program on a computer $N'$ that simulates your entire universe, and so in particular simulates  an instance of you who, in this
 simulation-within-a-simulation, is running a program on a  computer $N''$ that simulates your entire universe, and
so in particular ... Crucially, under the PCT, all those instances of you \textit{are} you; it is meaningless to ask which
of those instances ``is the real you'', with the others being ``just a copy''. This section ends with a
discussion of the philosophical quirks that would arise if the program being used to (self-)simulate universe
is encrypted, so that only the being with a special decryption key can understand the result of that computation.

The simulation and self-simulation lemmas allow us to define the ``simulation graph''. This is
the directed graph where each node is a universe 
containing a computer, and there is an edge from one node to another
if the (universe identified with the) first node can simulate the (universe identified with the)
second node. In \cref{sec:simulation_graph} I present a preliminary investigation of this graph.

Then in \cref{sec:elementary_properties} I discuss some of the mathematical properties that arise in both
simulation and self-simulation, in addition to those raised by consideration of the simulation graph. Specifically,
I use Rice's theorem to establish that many of the mathematical questions one might ask concerning simulation and self-simulation
are undecidable. 

Next in \cref{sec:math_implications} I discuss some of the very many open mathematical issues 
involving the simulation framework that I have not considered in this paper.

Finally, I begin \cref{sec:discussion} with a discussion of the implications of the results of this paper for arguments in
some of the earlier semi-formal work on the simulation hypothesis. After that I present 
quickly mention some of the ways that the paradigm implicitly considered in this paper involving the classical Church-Turing
thesis might be extended to apply to quantum and / or relativistic universes.

\section{Preliminaries}
\label{sec:preliminaries}

\subsection{Notation}
\label{sec:notation}

My notation is conventional. The set of of all positive integers is $\N$, and the set of non-negative integers is $\Z^+ = \N \cup \{0\}$.
 I write $|X|$ for the cardinality of any set $X$. 
In addition, for any set $X$ I write $X^*$ for the set of all finite strings of elements of $X$. Note that so
long as $X$ is finite, $X^*$ is countable. As an important example, $\B^*$ is the set of all finite bit strings. Since that set can be bijectively
mapped to $\N$, I will follow convention and treat finite bit strings as positive integers and vice-versa, with
the bijection implicit. 

As discussed in \cref{app:TMs}, I write
$T^m(x)$ for the (possibly partial) function given by running the TM with index $m$ on input $x$ until it halts (with the 
function undefined for input $x$ if it does not halt for that input). 
So $m$ ranges over all natural numbers $\N$.
%
Often I assume, implicitly or otherwise, that certain quantities are in the form of prefix-encoded bit strings.
In particular, I use some standard bijective encoding function of all tuples of bit strings into a single bit string, indicated using
angle brackets, $\anglebracket{., .}, \anglebracket{., ., .}$, etc. I assume that this encoding is non-decreasing in the number
of arguments, i.e., for any finite set of bit strings $\{b(1), b(2), \ldots, b(m)\}$, the length of
the bit string $\anglebracket{b(1), b(2), \ldots, b(m-1)}$ is not greater than that of  $\anglebracket{b(1), b(2), \ldots, b(m)}$.

Other important definitions and notations are in \cref{app:TMs} , e.g., of partial functions, and of
computable functions (here always assumed to be total). I also review the definition of ``computational universality''
in that appendix. In addition, there I review the definitions of a universal TM (UTM), and a prefix-free TM (the implementation of TMs I will often
be assume in this paper, implicitly or otherwise). I also define the instantaneous description (ID) of a 
Turing machine there. In the main body of this paper I will assume that the non-blank alphabet of the
TM, $\Lambda$, is just $\B$. So as described in \cref{app:TMs}, we can take the state of the tape(s) in any ID to be a finite string
in $\B^*$, even though strictly speaking the definition of TMs assumes infinitely long tapes.

Unfortunately (as happens all too often), there are some conflicts in the literature concerning terminology for
TMs. The reader should always check \cref{app:TMs} for the specific definitions used in this paper. Furthermore,
even if the reader is well-versed in TMs and the associated notation, and even if they are familiar with the recursion
theorem, they should still read \cref{app:recursion_theorem-rice_theorem}, since in this paper I will
use a slight extension of the recursion theorem, called the ``total recursion theorem''.


\subsection{Framework for analyzing CS theory of the simulation hypothesis}
\label{sec:framework}

To connect with the PCT and CS theory more generally, I will consider universes
that can be understood as evolving in discrete time, and that contain a subsystem 
that we will view as a ``computer''. Since I will want to take that computer to be computationally universal (and therefore
having exactly the computational power of a UTM),
I will assume that it initially has some arbitrarily large finite number of states,
where that state space can be enlarged dynamically, as needed, by an arbitrary amount. 

I will mostly be interested in cases where the computer simulates the evolution
of the state of the universe external
to the computer (the \textbf{environment} of the computer) and/or its own evolution. This means that the state space of
the environment must be finite (though arbitrarily large), to ensure that its state at a particular time can be
appropriately encoded on the input of the computer, i.e., of a UTM.

This leads me to write the state space of a \textbf{(computational) universe} as
\eq{
V = W \times N
} 
where the set $W$, representing the environment, is initially finite and cannot be extended dynamically.
On the other hand, $N$, representing
the state of (IDs of) the computer, is countably infinite (or what's equivalent for current purposes, $N$
is finite, but can be extended dynamically by arbitrary, finite amounts).

The elements of $W$ and $N$ are written as $w$ and $n$, respectively.
The elements of $V$ are written generically as $v = (v_W, v_N)$ or $(w, n)$. In general, the elements of $W$ as well as $N$ can
be indexed as multi-dimensional variables, e.g., as bit strings. However, I never need to make such
indexing explicit in this paper. Furthermore, sometimes it will be convenient to give subscripts to some of
these quantities, to indicate the time (which I take to be discrete). For example, I will sometimes
write $v_t = (w_t, n_t)$ for the time-$t$ state of the universe.

It will be necessary below to parameterize $N$ as $\underline{N} \times R$. 
In this decomposition, $R$ is a finite set that represents the state spaces of the internal variables of
the computer. So for example, in a conventionally represented TM, $R$ would be the state of the
TM's head, the position(s) of its pointer(s) on the
tape(s), etc.~\footnote{Strictly speaking, in a TM the pointer variable(s) can have any value in $\N$,
since they can point to an arbitrary position on the (countably infinite) tape(s). This technical concern can be addressed in
several ways. For example, we might have the states of the pointer variable(s) at any iteration \textit{after} the $k$'th
iteration given by the state of some special subsection of the tape(s), $\underline{n}$, with the states of the pointer variable(s) at earlier
iterations recorded in $R$. Another possibility would be simply to have $R$ be infinitely extendable. The precise
solution chosen is immaterial for this paper.}

$\underline{N} = \B^*$ is instead the state of the tape(s) in the case that the computer is
represented as a TM, or it is the state of the memory in the case that the computer is represented
as a RAM machine, or is represented as an appropriately modified laptop computer. 
The reason for writing
the countably infinite space $\underline{N}$ as the set of finite bit strings is so that we can physically implement it as
a system that initially has a finite state space, but whose state space can be expanded 
by an arbitrarily large finite amount an arbitrary number of times as the universe evolves. (See \cref{sec:example_V_our_universe}.)
I will write the elements of $\underline{N}$ as $\underline{n}$.


I will always assume that the initial, $t = 0$ state of $R$ is some special initialized value, $r^\varnothing$.
So $n_0$ is fully specified by $\underline{n}_0$, the initial state of the tape of the TM.
In addition, except when explicitly stated otherwise,
whenever I refer to the state of a computer for times  $t > 0$, I will only be interested
in the state of the tape at that time. Therefore except where explicitly stated otherwise,
I will treat $N$ and $\underline{N}$ as identical, with the associated individual states at iteration $t$
written as $n_t = \underline{n}_t$. (The major exception
to this convention occurs in \cref{sec:cheating_computers}, in which I need to consider $r_t$ for 
some times $t$ that are after initialization but before the TM has halted.)

I will informally use the term \textbf{physical universe} to refer to a special type of
computational universe. 
A physical universe is one that obeys one set of laws
of physics in all (of its) time and space, with one set of initial conditions, etc.\footnote{For simplicity, I sidestep the issue of multiverses 
with different physical constants but otherwise identical physical laws, e.g., arising
from a shared inflation epoch.}
As a special case, I use the term \textbf{cosmological universe} to refer to an
entire physical universe, containing all of its relevant variables -- t 
just like our physical universe.

In general, the theorems derived below allow $V$ and $V'$ to be (portions of) different cosmological universes, obeying different
laws of physics. They could also be sub-regions of the same cosmological universe though (in which case they obey the
same laws of physics). One way this could occur
is if they occupy physically distinct regions of the same physical universe. (See \cref{sec:example_V_our_universe}.)
This case of multiple computational universes in the same cosmological universe
will be important in the discussion of the simulation graph below in \cref{sec:simulation_graph}.

To simplify the analysis, I do not assume that the models of  physical
universes that I investigate in this paper capture those universes \textit{in toto}, e.g., if the state spaces of those entire physical
universes are in fact uncountably infinite. (So unless explicitly stated otherwise, I do not have in mind a 
cosmological universe when I refer to a physical universe.) Similarly, I will not assume that a simulation running in a computer inside a universe
models the dynamics of an entire physical universe. Rather throughout this paper the expression ``universe'' should be understand
as shorthand for ``possibly coarse-grained subset of a universe''. However I do assume
that the physical universes obey deterministic dynamics, be it over the original physical state space (assuming it is countable)
or some coarse-grained version of such a state space. As an example, in \cref{sec:example_V_our_universe} 
I present a detailed example of what such a ``coarse-grained subset of a universe'' could be for our particular physical universe.

I write the evolution of the universe from an initial, time-$0$ state $(w_0, n_0)$ to a time $t$ state as
a vector-valued \textbf{evolution function} $g$ of the initial state of that universe, 
\eq{
g(t, w_0, n_0) = (w_t, n_t)
\label{eq:g}
}
Note that the image of $g$ is the Cartesian product $W \times N = W \times (\underline{N} \times R)$.
I use usual notation for components of vector-valued functions. So in particular, $g(t, w_0, n_0)_{\underline{N}}$ is the 
$\underline{N}$-component of $g(t, w_0, n_0)$. 

Unless specifically state otherwise, from now on I restrict attention to universes
whose evolution function $g$ is computable (see \cref{app:TMs}). Note that
for ``computable'' to even be a potential property of $g$ means that I am implicitly assuming that
the outputs of $g$ are actually single bit strings (or equivalently, single counting numbers) that encode a
pair of bit strings, as in \cref{eq:g}.

\subsection{What it means for one universe to simulate another}

The term ``simulation'' was not given a formal definition in any of the previous literature on the simulation hypothesis.
Moreover, ``simulation'' (and the associated term ``bisimulation'') already has a formal definition in the CS theory of state transition systems~\cite{wikipedia_state_transition_system}.
However, despite its name, this definition from CS theory does not
describe what ``simulation'' is loosely understood to mean in the context of the simulation hypothesis.

In this paper $V$ ``simulates'' $V'$ if for all initial states $v'_0$ of $V'$, and all amounts of
time $\Delta t'$ into the future of $V'$ that we might want to simulate the dynamics of $V'$, there
is some associated initial state of the simulating universe $V = N \times W$, together with a time into the future of
the simulating universe, $\Delta t$, such that the state of the computer $N$ at that future time
$\Delta t$ gives us the desired state of $V'$ at \textit{its} future time $\Delta t'$.

I formalize this using an appropriate prefix code $\langle ., . \rangle$ as follows:
\begin{definition}
A universe $V =  W \times N$ with evolution function $g$ \textbf{simulates} the evolution of 
a universe  $V' =   W' \times N'$ with evolution function $g'$ iff there exist three functions
\eq{
&\TT(\Delta t', w'_0, n'_0) \in \N \nonumber \\
&\WW(\Delta t', w'_0, n'_0) \in W \nonumber \\
& \NN(\Delta t', w'_0, n'_0) \in N \nonumber
}
such that for all $\Delta t' \in \N, w'_0 \in W', n'_0 \in N'$, $t \ge \tau$,
\eq{
g(t, \omega, \eta)_{\underline{N}} = \langle g'(\Delta t', w'_0, n'_0) \rangle  
\label{eq:simulate_def}
}
where as shorthand,
$\tau :=  \TT(\Delta t', w'_0, n'_0), \omega := \WW(\Delta t', w'_0, n'_0), \eta :=  \NN(\Delta t', w'_0, n'_0)$.
\label{def:sim}
\end{definition} 
\noindent Note that the LHS of \cref{eq:simulate_def} is the second of the two components of the vector-valued function $g$, 
while the RHS is an encoding of both components of $g'$ into a single variable. 
Note also the requirement in \cref{def:sim} that \cref{eq:simulate_def} hold for \textit{all} $t \ge \tau$, and so the 
state of the simulating computer $N$ does not change after it completes its simulation of the future state of $V'$. 
This just means that I require that the simulating computer halts when it completes its simulation. I also require that the ID of the computer $N$ be
replaced \textit{in toto} (i.e., uniquely, with all other variables fixed to some predefined values) 
by the output of its simulation program. So for example, if $N$ is a multi-tape prefix TM, this
means that when $N$ halts with the result of its simulation on its output tape, 
all the other tapes --- the intermediate work tapes and the input tape --- have been re-initialized to be all blanks.


I will sometimes say that $V$ ``can simulate'' $V'$ rather than say that it ``simulates'' $V'$.
I also sometimes say that $V$ simulates $V'$ \textbf{for simulation functions} $\TT(.,.,.)$, $\WW(.,.,.)$, and $\NN(.,.,.)$
if \cref{def:sim} holds for that particular triple of functions. 
In addition I say that $V$ {\textbf{computably}} simulates $V'$ if it simulates $V'$, and in addition the three functions 
$\TT(\Delta t', w'_0, n'_0), \WW(\Delta t', w'_0, n'_0), \NN(\Delta t', w'_0, n'_0)$ are all computable. Unless specified otherwise,
whenever I refer to ``simulation'' in this paper I implicitly assume it is computable.

Note that  \cref{def:sim} does not require that 
the simulating computer $N$ calculates the future
state of the universe $V'$ by itself, independently of the initial state of the environment $W$ outside of $N$, i.e., independently of the
value of $w_0$. (Formally, the LHS of \cref{def:sim} is not independent of $\omega$, which depends on the initial
state of $W$.) The reason for this flexibility is to allow the computer $N$ to retrieve
the specific information it needs to perform its simulation of the dynamics of the specific state
$v'_0$ from some putative super-aliens who are running that computer $N$, and who exist in $W$, the environment of $N$ (in the sense that
the precise state of those aliens is specified by the elements of $W$). Note though that this
freedom also allows the beings running the simulation computer $N$ to intervene on the dynamics of
that computer at any time they want, e.g., by overwriting the simulation program being run in the computer. They can
even ``pull the plug early'' on that simulating computer, before it finishes its computation.

Note also that  \cref{def:sim} allows the initial state of the simulating computer, $n_0$, to vary 
if we vary the time into the future, $\Delta t'$, that the simulating
computer is calculating. Indeed, the definition allows $n_0$ to vary for different $\Delta t'$ even if $N$ is simulating
the future state of $V'$ for all those values of $\Delta t'$ evolving from the same initial state $v'_0 \in V'$.
Concretely, I, a super-alien, might use one program to compute the state $v'$ of a universe $V'$ at the time
$\Delta t'_1$ into the future of that universe, and use a different program to compute the state of $V'$ at a time 
$\Delta t'_2$ into the future of that universe.
(However, this flexibility is circumscribed if we restrict attention to ``time-consistent'' universes, as discussed 
below in \cref{sec:cheating_computers}.)

\cref{def:sim} allows both $w_0$ and $n_0$ to vary if we change $w'_0$, even if $n'_0$ is fixed.
Often we are interested in a more restrictive notion of simulation, where for a fixed $n'_0$, 
changing $w'_0$ does not change $n_0$ (even though changing $w'_0$ is allowed to change $w_0$ in general). 
Intuitively, this restrictive form of simulation corresponds to the case where the simulation program is fixed, 
and reads in the precise initial state of the system it is going to simulate after it starts running, from
some appropriate variable in that simulation program's external environment. As an example, this
form of simulation would be met if
$N$ were a UTM, so that $n_0$ specifies the precise TM that $N$ is implementing,
while $w'_0$ is extra information that is subsequently ``read in'' by that TM $n_0$ from its
external universe $w_0$, after $n_0$ starts running. 


We can formalize this restricted form of simulation with a simple extension of \cref{def:sim}.

\begin{definition}
Suppose that $V$ simulates the evolution of $V'$ for three functions $\TT, \WW, \NN$. Then $V$ \textbf{freely simulates} 
the evolution of $V'$ if for some specific fixed $n'_0$, $\NN(\Delta t', w'_0, n'_0)$ is independent of $w'_0$.
%
\end{definition}

Finally, the definitions above only stipulate that the simulating computer $N$ 
eventually outputs $v_{\Delta t'}$, the future state of $V'$ at one specific time, ${\Delta t'}$. It is straight-forward to
extend these definitions to have $N$ output an entire trajectory of $L$ such future states instead. To do
this we would replace the first arguments of $\TT, \WW$ and $\NN$ with a vector
$\vec{\Delta t'} \in \N^L$. We would also extend the definition of the evolution function, to have
\eq{
g'(\vec{\Delta t'}, w'_0, n'_0)
}
be the states of $V'$ at the sequence of times $\vec{\Delta t'}$ when it starts at time $0$ with the state $(w'_0, n'_0)$.
We would make an analogous extension to the definition of the evolution function $g$ of the simulating computer's universe.
Finally, we would modify the condition in \cref{eq:simulate_def} to say that $V$ \textbf{trajectory-simulates} $V'$ if

$ $

\indent For all $\vec{\Delta t'} \in \N^L, w'_0 \in W', n'_0 \in N'$, $t \ge \tau$,
\eq{
g(t, \omega, \eta)_{\underline{N}^L} = \langle g'(\vec{\Delta t'}, w'_0, n'_0) \rangle  
\label{eq:simulate_def_mod}
}
\indent where $\tau :=  \TT(\vec{\Delta t'}, w'_0, n'_0), \omega := \WW(\vec{\Delta t'}, w'_0, n'_0), \eta :=  \NN(\vec{\Delta t'}, w'_0, n'_0)$.

$ $ 

\noindent For simplicity in this paper I do not consider trajectory-simulation, focusing on single-moment simulation. 
However, all the results below still apply for trajectory-simulation, with minor terminological changes.

%
%

To minimize notation, in the sequel I will implicitly
choose units of physical time so that under the dynamics of any universe I am considering, the physical computer
in that universe takes one unit of (physical)
time to run one iteration of the computational machine it is implementing.
(For example, if that computational machine is a UTM, then each iteration of the UTM takes one unit of physical time.)

\subsection{The Physical Church-Turing Thesis}
\label{sec:PCT}

Even restricting consideration to computers that can be described using
classical physics, there are many different semi-formal definitions of the PCT in the literature~\cite{piccinini2010computation,piccinini2011physical,copeland2023church,copeland2018church,aaronson2013philosophers}.
If we extend consideration to include quantum computers~\cite{nielsen2010quantum}, there are even more 
definitions~\cite{arrighi2012physical,nielsen1997computable}. 

Whether in fact our particular cosmological universe obeys the PCT, be it the classical or quantum PCT, has been subject to endless argument~\cite{aaronson2013quantum,parberry2013knowledge,arrighi2012physical,piccinini2011physical,nielsen2006quantum}. 
In related work, some researchers have designed purely theoretical, contrived physical systems that are uncomputable in some
sense or other~\cite{pour1982noncomputability,cubitt2015undecidability,shiraishi2021undecidability} (see also~\cite{chaitin2011goedel}). 
This work has resulted in attempts to define the PCT to exclude the case of physical systems whose future is uncomputable
but which cannot be constructed by we humans in a finite amount of time. This amounts to tightening the PCT to concern not just what
systems can be simulated, but rather what systems can be constructed and then simulated.

In any case, as mentioned above, for the purposes of this paper, it does not matter whether some 
particular form of the PCT applies to \textit{our} specific universe. What matters is the CS theory implications
of  universes simulating other universes, and in particular the implications if the PCT holds for such universes.
Accordingly, for current purposes, I make the following (fully formal) definition:
\begin{definition}
The {\textbf{Physical Church-Turing thesis (PCT) holds for universe $V$}} iff the evolution
function $g(., ., .)$ of $V$ is computable. 
\label{def:PCT}
\end{definition} 
\noindent (It would seem that this is the first fully general definition of the PCT, being applicable even to universes whose laws of
physics differ from ours.) By \cref{def:PCT}, if the PCT holds for $V$, there must be a UTM that (halts and) outputs the
vector value of $g(., ., .)$, for all values of $g$'s arguments. (More precisely, there must be such a UTM
that outputs the string $\anglebracket{g(., ., .)}$ for all values of $g$'s arguments if it receives the encoded
version of those arguments as its input.)

All the analysis in this paper will assume that at least some of the universes being discussed obey the PCT.
This means that the analysis below does not apply to any universes so many levels of
computational power above our own that they can contain computers capable of super-Turing computation~\cite{aaronson2013philosophers}.
In particular, the analysis would not apply to any such universes that are simulating our universe.

Much of the earlier literature on the ``physical Church-Turing thesis'' accords a prominent
role to humans, and their abilities (or lack thereof), e.g., in arbitrarily configuring the initial
state of physical systems, or in observing their subsequent physical state. There is no
such role in \cref{def:PCT}. All the PCT means in this paper is that evolving the computational
universe $V$ does not require a computational machine more powerful than a TM. In fact,
the definition would be satisfied even if that evolution can be calculated on a machine that is strictly weaker than a TM,
e.g., a finite automaton. Furthermore, if $V$ is a sub-region of some cosmological universe that obeys the PCT
as defined in~\cref{def:PCT},
it could be that machines more powerful than a TM are required to calculate the evolution
of some other sub-region of that cosmological universe, different from $V$.

For these kinds of reasons, some readers might argue that \cref{def:PCT}
doesn't exactly capture all of the various properties that have been referred to
as the ``physical Church Turing thesis'' in the literature.
In some senses \cref{def:PCT} is more like a definition of one of the related concepts inspired by modern physics, e.g., some forms of
ontic structural realism~\cite{french2010defence,mccabe2005possible,mccabe2006structural,ainsworth2010ontic}
or the level IV multiverse~\cite{tegmark1998theory,tegmark2008mathematical}. 
But for current purposes, it suffices to simply accept~\cref{def:PCT}, and ignore any objection
such terminology might raise in certain researchers. 

Finally, note that the set of spaces $W \times N$ is countably infinite, if we restrict that set to contain all
finite spaces $W$. Therefore the set of universes defined by the specification of such a space, together with an evolution function
that obeys the PCT, is also countably infinite. Moreover, many of the considerations of the ``simulation
hypothesis'' in the literature implicitly assume such a universe and evolution function.
Finally, note that it is impossible to assign a uniform probability  
distribution to the set $\N$. This establishes the claim made in the introduction, that it is impossible to 
assign a uniform probability distribution to the kind of universes often considered in the literature on 
the simulation hypothesis.

\subsection{The Reverse Physical Church-Turing Thesis}
\label{sec:RPCT}

Loosely speaking, the PCT says that the dynamics of any universe that we are considering 
can be computed on a UTM.
One can ``reverse'' the requirement that a universe obey the PCT, which results in the requirement that a universe 
contains a UTM in it. If it obeys such a reversed PCT, a universe could implement all TMs. (Note that this
does not mean that the universe itself is a universal TM; it could have super-Turing power, in which case it is
not any type of TM at all.)

One might imagine that the reversed PCT could be formalized in a way that among other things requires
that a universe's computer $N$ evolves independently of the rest of that universe.
However, in general that cannot be assumed in the analysis below --- we will need to allow the beings running the computer
to provide information to that computer after it starts running.
(As an example, this is the case whenever one runs a computer program that gets input from the user or an external
disk drive after it starts running.) This means that that computer $N$ does not evolve autonomously as a UTM. 
So we cannot impose this simple version of reverse PCT. 

On the other hand, we can require that $N$ {\textit{effectively}} implements a UTM. This is done as follows:
\begin{definition}
The \textbf{reverse physical Church-Turing (RPCT) holds for universe $V =  W \times N$} with evolution function $g$ iff
there exist three functions 
\eq{
&\widehat{\TT}(k, y) \in \N \nonumber \\
&\widehat{\WW}(k, y) \in W \nonumber \\
&\widehat{\NN}(k, y) \in N \nonumber
}
that are
well-defined for all TM indices $k$ and all $y \in \B^*$ such that $T^k$ halts on input string $y$, where
for all such $(k, y)$,
\eq{
g\left(\widehat{\TT}(k, y), \widehat{\WW}(k, y), \widehat{\NN}(k, y)\right)_{\underline{N}} = T^k(y) \nonumber
}
\label{def:RPCT}
\end{definition}
\noindent I say that three functions $\widehat{\TT}, \widehat{\WW}, \widehat{\NN}$ all taking 
$\N \times \N \rightarrow \N$ \textbf{have the RPCT properties for $g$} if they obey the properties listed 
in \cref{def:RPCT}. Abusing terminology, I also say that the {\textbf{computable} RPCT holds for $V$
iff all three functions $\widehat{\TT}(k, y),  \widehat{\WW}(k, y), \widehat{\NN}(k, y)$ are computable for all 
pairs of arguments $(k, y)$ such that $T^k(y)$ halts. (They may be undefined for other $(k, y)$, i.e., they may
be only partially computable.) Unless specified otherwise,
throughout this paper I will assume that whenever the RPCT holds, it's computable.

Broadly speaking, the RPCT says that the system $N$ operates like a UTM for all pairs $(k, y)$ such that $T^k(y)$ is defined,
where $k$ and / or $y$ may be encoded in some degrees of freedom in $w_0$ rather than directly in $n_0$. (That 
freedom to have $k$ and / or $y$ specified in the environment of the computer $N$
allows that computer to do things like observe its environment
to retrieve the input string for a computation from its environment before running that computation.)  

I say that the \textbf{pristine} RPCT holds if for all $k, y$, the RPCT holds with 
 \eq{
\label{eq:7}
\widehat{\NN}(k, y) &= k \\
\widehat{\WW}(k, y) &= y
\label{eq:8a}
}
%
\cref{eq:7,eq:8a} mean that
the physical system implementing the UTM is initialized with the precise program that it is
to supposed to implement, but not the actual data that program will be run on. In general, that data (the value of $y$)
would be transferred into the computer $N$ at a subsequent iteration, after this initialization of the UTM,
but before the UTM starts to run. As an example, this is what would occur if we were to initialize
a laptop with a universe-simulating program,
with the precise data that program is to run on is fed into the program before the laptop starts evaluating that simulation. 
(See \cref{sec:example_V_our_universe}.) 

Note that in the definition
of the pristine RPCT I don't introduce
the extra notation that would be needed if the extra information concerning $y$ that $N$ would use is contained
in a proper subset of the variables in $W$. Instead I simply equate the value $y$ with $w \in W$ in its entirety.
This is just for pedagogical simplicity.
%
In any case, when assuming the RPCT below I will not implicitly assume that the \textit{pristine }RPCT holds 
unless I explicitly say so.

Just as the PCT as defined in \cref{def:PCT} has no role for humans, the RPCT has no role
for them. In particular, \cref{def:RPCT} does \textit{not} say that humans could configure the
initial state of $V$ to implement the dynamics of any desired TM. It simply says that there is some such initial
state of $V$ that could implement that dynamics.

The RPCT is accepted by many researchers, implicitly or otherwise. (Indeed, it is commonly
confounded with the PCT.) For example,
Scott Aaronson wrote in a blog post on Feb. 8, 2024 that 
``My personal belief is that... `yes,'  in some sense (which needs to be spelled out carefully!)
 the physical universe really is a giant Turing machine.'' See
also~\cite{miranda-cardona2021constructing,moore1990unpredictability,pour1982noncomputability}
and related literature for more formal considerations of (what amounts to) the RPCT.

Despite the popularity of the RPCT among researchers, even if the PCT holds 
in our specific universe, that does not mean that the RPCT has to hold as well. In particular, cosmological considerations
could prevent it from holding~\cite{lloyd2013universe}.
So the results below specific to universes that obey the PCT and / or universes that obey
the RPCT might not apply to our specific universe. However, approximations of the analysis might
apply even if the PCT and/or RPCT don't hold exactly in our universe, depending on how precisely 
those properties are violated.

Even without worrying about the laws of physics in our universe,
one might suppose that the RPCT is logically impossible in our cosmological universe,
if the PCT holds for our universe.
After all, the RPCT requires that the set $X$ of physical variables of a universe decomposes as
$X = W \times N$ where $N$ is a physical system that (once it reads in the data $y$) implements a UTM.
In other words, a proper subset of the spatial degrees of freedom of the
universe constitutes a physical structure $N$ capable of
implementing any TM. But the PCT thesis supposes that such an $N$ would be able to simulate
the dynamics of \textit{all} of $X$. So this $N$
would have to be able to simulate \textit{itself}, {at the same time} as it is also simulating
the entire rest of $X$, outside of $N$. This would seem to imply a contradiction,
that $N$ is more computationally powerful than $N$ is. 

If this argument were valid, the RPCT would be impossible.
And so in particular, no matter what the actual laws of physics,
it would not be possible for us to be part of a simulation by a computer, if that computer were
itself contained in our cosmological universe. In other words, if this argument were valid, it would
be a proof that the simulation hypothesis must be wrong.

One might be suspicious of this argument though, since it is quite similar to the arguments that were made 
in the last century that
no physical system can make an extra copy of itself without destroying itself. 
(These arguments that copying required destruction of the original were used to make the case that
the common definition of life involving replication must be wrong, or at least deficient.) Responding to these arguments, Von Neumann 
designed his ``universal constructor'' in a cellular automaton setting. This demonstrated explicitly that it \textit{is} possible
for a system to copy itself without harming itself in the process.\footnote{It is interesting to note that 
Von Neumann's proof of this property of his universal constructor is essentially identical
to Kleene's earlier proof of his second recursion theorem --- and that it is hard to imagine that Von Neumann did not
know of Kleene's earlier result, despite not citing it in his work.} In modern parlance, computer viruses are
mathematically possible -- which is perhaps comforting, since in fact they exist.

\subsection{Sketch of how a portion of our cosmological universe could obey the PCT and RPCT}
\label{sec:example_V_our_universe}

In the rest of this subsection I sketch how the framework defined above might apply to a portion
of our actual cosmological universe, evolving according to the laws of physics as we currently
understand them. 

Choose some particular inertial frame to describe physical systems in our universe.
(Sticking to this specific inertial frame will allow us to circumvent the complications of special relativity.)
Let $Y$ be a special subset of the space-time of our cosmological universe defined by a cylinder in our
chosen inertial frame, a cylinder whose base
consists of a $3$-dimensional region, $Y_0$. $Y_0$ contains an infinite number of accessible degrees
of freedom, and all of the points in $Y_0$ have the same value of time in our chosen inertial frame, $t_0$,
a time which is long after the Big Bang. (Having $t_0$ be long after the Big Bang 
will allow us to circumvent the complications of cosmological issues involving general relativity and
the boundary conditions of our cosmological universe, and will
allow us to talk about initializing various physical variables in a computer that resides
in $Y$.) For completeness, suppose that $Y$ extends infinitely to the future of time $t_0$. 
Discretize time (in our co-moving inertial frame) going forward from $t_0$, writing the
time at the end of each interval as $t_i$, $i \in \N$, and the associated state of our region as $Y_i := Y_{t_i}$.

Let $V = W \times N$ be a coarse-graining of that $3$-dimensional region defining $Y$ into a set of classical (non-quantum) bins, 
where $W$ is finite and ${N}$ is countably infinite. As described above, it is useful to parameterize ${N}$ as
the set $\B^*$ of all finite strings of bits.
Suppose that the variables given by that coarse-graining follow deterministic evolution.
This requires in particular that $V$ be physically isolated from the rest of the universe, and
so they collectively obey Hamilton's equations. Note
that by considering the dynamics of $V$, we can circumvent
the complications of quantum mechanics, and in particular of extending the PCT thesis to
the quantum realm. 

Since Hamilton's equations are computable, and $X_t$ obeys Hamilton's equations,
the PCT is obeyed by the dynamics of $X_t$. We also assume the projected dynamics down to $N$ implements a UTM (though perhaps
it does so slowly, limited by the speed of light among other factors). Therefore the RPCT
is also obeyed by the dynamics of this universe, $V$.

Given this general setup, choose $w_0 \in W$ arbitrarily. Also
pick an initial ID $n_0 \in N$ of the UTM that has an arbitrary (blank-delimited) finite string on the input tape,
with all other tapes (if any) initially fully blank, and the pointers of the UTM in their initialized positions.

Choose the dynamics of the entire universe $V$ so that  the
first thing that happens when it starts evolving from its initial state is that 
a copy of $w_0$ is placed after the string $n_0$ on the input tape of the 
(physical system implementing) the UTM. Physically, this could reflect the process
in which the being who is going to run the simulation program on the
computer feeds that computer the input data it needs concerning
the external universe. Alternatively, it could reflect the computer being coupled
to an observation device which observes that data and provides it to the
computer. After that the dynamical laws  (Hamilton's equations in this universe) 
evolve both $W$ and $N$, perhaps via a coupled dynamics, or perhaps completely independently of one another.

As an aside, it is unusual in standard physics to consider state spaces 
$\B^*$ (which is the space we assign to $N$). 
However, such spaces arise naturally when modeling a discrete computer, e.g., a laptop, whose memory is always finite --- but
can be expanded by an arbitrary finite amount an arbitrary number of times. Viewing that memory
as $M$, we can formalize its dynamic expansion in
terms of dynamics over the uncountably infinite space $\B^\infty$, by ``embedding'' $N$ into $\B^\infty$ as the indexed set of spaces
$N_t = \B^{K(t)}$ for some function $K : \Z^+ \rightarrow \N$. We define this function $K(t)$
recursively. First, specify $k(0)$ to be some arbitrary finite counting number. 
Then if at time $t$ there are any bits in $\B^\infty \setminus N_t$ whose state at $t+1$ is  
causally dependent on the state of $n_t \in N_t$, $K(t)$ increases enough to include those bits. 
(Note the implicit assumption that at every iteration $t$, there are only a finite number of such bits whose
state may depend on $n_t$.)

See \cref{app:CAs_not_TMs} for discussion of some subtleties of the scheme outlined above for implementing a universe $V$ in
our actual, cosmological universe.

\section{The simulation lemma }
\label{sec:sim-lemma}

In this section I first derive the simulation lemma. I then discuss some of its features,
and in particular why it might seem to imply that self-simulation is impossible.

\subsection{Proof of the simulation lemma}

In this subsection I first state the simulation lemma formally and then prove it.

\begin{lemma}
If a universe $V =  W \times N$ obeys the RPCT, then it can simulate any universe
$V'$ that obeys the PCT.
\label{lemma:sim}
\end{lemma}

\begin{proof}
By \cref{def:PCT}, since $V'$ obeys the PCT, its evolution function $g'$ is computable. Therefore there is an index $k \in \N$ such
that 
\eq{
T^k(\langle \Delta t', w', n'\rangle) = \langle g'(\Delta t', w', n')\rangle \nonumber
}
for all $\Delta t' \in \N, w' \in W', n' \in N'$. Therefore by  \cref{def:RPCT}, since $V$ obeys the RPCT,
there are three functions $\widehat{\TT},  \widehat{\WW},  \widehat{\NN}$ such that
for any triple $(\Delta t', w', n')$ and associated finite string $y := \langle \Delta t', w', n'\rangle$, 
\eq{
g\left(\widehat{\TT}(k, y), \widehat{\WW}(k, y), \widehat{\NN}(k, y)\right)_{\underline{N}} &= T^k(y) \nonumber \\
	&=  \langle g'(\Delta t', w', n') \rangle 
\label{eq:sim_lemma_1}
} 
Next define
\eq{
\TT(\Delta t', w', n') &:= \widehat{\TT}(k, \langle \Delta t', w', n'\rangle) \\
\WW(\Delta t', w', n') &:= \widehat{\WW}(k, \langle \Delta t', w', n'\rangle) \\
\NN(\Delta t', w', n') &:= \widehat{\NN}(k, \langle \Delta t', w', n'\rangle) 
\label{eq:9''}
}
Plugging these three definitions into \cref{eq:sim_lemma_1} and then comparing to \cref{def:sim} completes the proof.
\end{proof}

I will refer to \cref{lemma:sim} as the \textbf{simulation lemma}.

\subsection{Why the simulation lemma might seem to preclude self-simulation}
\label{sec:why_cant_self_sim}

The simulation lemma tells us that if there are some super-sophisticated aliens in a universe
$V$ in which the RPCT holds, and if our universe obeys the PCT, then it's possible that we
are simulations in a computer that the aliens are running.
(Of course, we would never know it if that were the case.) On the other hand,
suppose that, as many have argued, the PCT does \textit{not} hold in our universe, because our universe's evolution
cannot be evaluated on a TM. In such a case, even if the RPCT holds in the universe that the aliens
inhabit, there is no guarantee that we may be in a simulation that they are running. This formalizes the intuitive idea
that our universe has to be ``sufficiently simple, computationally speaking'' in order for us to be simulations in a computer
of some super-sophisticated aliens.

The situation becomes more philosophically interesting if we consider the case that the universe
being simulated, $V'$, obeys the PCT, and in addition \textit{is $V$ itself}.
This would mean in particular that $N$ is being used to simulate the evolution of $N$.

To investigate this possibility, note that even if the conditions in the simulation lemma hold, that lemma doesn't in some sense specify what
argument $(\Delta t', w'_0, n'_0)$ to simulate. In particular, it does not tell us what value $n'_0$ the computer $N'$
that $N$ is simulating would need to start with in order for that simulation to give the dynamics {of
$V$ itself}. To understand the implications of this, suppose that: i) the  RPCT
holds for our universe, so we have a computer $N$ we can run to simulate the evolution of any universe 
that obeys the PCT; ii) our universe itself obeys the PCT.  \cref{lemma:sim} would seem
to imply that when those two conditions are met, we can
run a simulation of our own universe, including ourselves. The issue  is that given
our definition of ``simulate'', we need to specify the precise initial state of the universe
being simulated. The simulation lemma provides no means for us to choose that
initial state for $N$ to run a simulation of $V$, nor to initialize our computer $N$ in order to 
run a simulation of ourselves. It provides no way to have a universe $V$ determine how to simulate \textit{itself}.

In fact, if we take $V = V'$ in the proof of \cref{lemma:sim}, then \cref{eq:9''} becomes a fixed point equation:
\eq{
n' &:= \widehat{\NN}(k, \langle \Delta t', w', n'\rangle) 
\label{eq:9'}
} 
We have no guarantee that the solution $n'$ to this fixed point equation is computable. In fact, \textit{a priori}, one might
suspect that it is possible for there not to be any solution to \cref{eq:9'} whatsoever, computable or otherwise. 
In light of \cref{lemma:sim}, if that were the case, it would mean that the conditions for the simulation lemma cannot be met, 
i.e., that it is not possible for a universe $V$ to obey both the PCT and the RPCT.
Indeed, none of the analysis above establishes that it is logically possible for a universe to obey both
the PCT and the RPCT. That means that it may be mathematically impossible to meet the conditions for \cref{lemma:2}.

An associated concern is that \cref{lemma:sim} only establishes the possibility of $V$ simulating $V'$. It does
not establish the possibility of \textit{free}
simulation. So even if we can establish that there is a solution to the fixed-point equation \cref{eq:9'}, there might
only be one, i.e., there might only be a very specific initial state of our own universe that we can simulate. 

In fact, if a computer were to simulate the evolution of itself up to a certain time in the future, 
that means in particular that it would simulate a ``copy'' of itself running a computer
which is simulating of a copy of itself running a computer which is simulating, etc., \textit{ad infinitum}. 
In other words, one might expect that self-simulation
would require an infinite regress of computers within simulations of computers. That in turn would suggest
that the computer could never complete such a simulation of itself in finite time (assuming that the computer
operates at a finite physical speed).


In the next section I prove that these concerns do not in fact prevent a universe from
simulating itself. In fact, not only is there a solution to \cref{eq:9'}, and not
only is there an explicit algorithm to construct that solution, but that solution is computable, i.e., the algorithm
that computes it is guaranteed to halt if implemented on a Turing machine.
This means that it \textit{is} possible for a universe to obey both the PCT and the RPCT. More
interestingly, it means that we can construct a computer such that for any given finite time interval $\Delta t$ and initial state of our universe
outside of that computer, we can load a program onto that computer which is guaranteed to
halt in finite time after having simulated the full state of our universe at the time $\Delta t$ into
the future --- including in particular the state of that computer itself at that time. 
It also means that we could be inside a simulation being run on a computer that we ourselves are running
(supposing our universe obeys both the PCT and RPCT of course).

\section{The self-simulation lemma}
\label{sec:self-sim-lemma}

In this section I begin by proving that despite the arguments in \cref{sec:why_cant_self_sim}, in fact self-simulation \textit{is}
possible. 
The key is to have a time delay between the
future moment for which the state of the universe is being simulated on the one hand, and when the computer simulation of 
that future state completes on the other hand.

After presenting that proof, I discuss certain important features of self-simulation. I end this section
by discussing  how 
the possibility of self-simulation differs from various forms of ``self-reference'' considered in the literature.

\subsection{Proof of the self-simulation lemma}

The proof of the simulation lemma just relies on elementary properties of TMs, and the assumptions
that the PCT and RPCT both are obeyed. We need more than
that to establish the possibility of free self-simulation. Specifically, we need to also
use the total recursion theorem, and we need to strengthen the assumption that the RPCT holds into
assuming that the pristine RPCT holds.
\begin{lemma}
If both the PCT and the pristine RPCT hold for a universe $V = W \times N$, 
then for all $\Delta t$ there exists $n_0 \in N$ such that $V$ freely simulates $V$ for $n_0, \Delta t$.
\label{lemma:2}
\end{lemma}
}

\begin{proof}
Since the PCT holds for $V$, its evolution function $g(., ., .)$ is computable (not just partial computable). 
Therefore if we fix $\Delta t$, the first argument of $g(., ., .)$,
and then invoke the total recursion theorem, we see that there exists $n^*$ such that 
\eq{
T^{n^*}(w_0 ) 	&=  g( {\Delta t}, w_0, n^*)
\label{eq:1aa}
}
for all $w_0 \in W$,
where both $g(\Delta t, w_0, n^*)$ and $T^{n^*}(w_0)$ halt for all inputs $w_0$.
Note that the TM index $n^*$ will depend on both $g$ and $\Delta t$ in general.

Now apply our assumption that
the pristine RPCT holds for $V$ to the LHS of \cref{eq:1aa} and then plug in the RHS of that
equation. This shows that there must exist a computable function
$\widehat{\TT}$ such that for all $w_0 \in W$, 
\eq{
g\left(\widehat{\TT}(n^*, w_0), w_0, n^*\right)_{\underline{N}} &= T^{n^*}(w_0)
\\
&= g(  {\Delta t}, w_0, n^*)
\label{eq:16}
}

Finally, if we now define
\eq{
\TT\left(\Delta t', w'_0, n'_0\right) &:= \widehat{\TT} \left(n'_0, w_0\right) \\
\WW\left(\Delta t', w'_0, n'_0\right) &= w'_0 \\
\NN\left(\Delta t', w'_0, n'_0\right) &= n'_0
}
for all $\Delta t', w'_0, n'_0$, then \cref{eq:16} can be re-expressed as
\eq{
g(\TT\left(\Delta t', w'_0, n^*\right), \WW\left(\Delta t', w'_0, n^*\right), \NN\left(\Delta t', w'_0, n^*\right)) &= g(  {\Delta t}, w_0, n^*)
}
Plugging this into \cref{def:sim} completes the proof that $V$ simulates the evolution of $V$ for $n_0 = n^*, \Delta t$. Since
\cref{eq:16} holds for all $w_0$, while $n_0$ is fixed to $n^*$,
we see that in fact $V$ freely simulates the evolution of $V$ for $n^*, \Delta t$.
\end{proof}


I will refer to \cref{lemma:2} as the \textbf{self-simulation lemma}. It says that for any fixed $\Delta t$, there is an associated initial state
of the computer such that for any initial state of the rest of the universe, $w_0$, that computer is guaranteed to halt and to output
the state of the entire universe at time $\Delta t$.

\subsection{Important features of the self-simulation lemma}
\label{sec:features_self_sim}

Recall the convention that ``simulate'' implicitly means ``computably simulate''. So
the self-simulation lemma not only says that there is an initial state of the computer, $n_0$, for which the computer simulates the entire
universe including itself. It says that $n_0$ is a computable function. Specifically, that value $n_0$ is the solution to \cref{eq:1aa},
and so implicitly depends on the combination of the time into the future that we want to simulate, $\Delta t$, and the 
evolution function $g$ (which is encoded as the index of a TM). 

By definition, that means there is a halting program that
constructs that $n_0$. So we can run that program (on any UTM we like, whether a physical system in $V$ or
not) and be assured that it will eventually finish and tell us what \textit{other} program $n_0$ to
load into our computer $N$ in order to simulate the evolution of our universe.

The evolution function $g$ arising in the self-simulation lemma can be encoded as  an integer (since it is a computable function,
by hypothesis). Moreover, there is an implicit function given by the self-simulation lemma
that takes the combination of $\Delta t$ and $g$ to the initial state of the self-simulating computer.
(See comment just below \cref{eq:1aa} in the proof of \cref{lemma:2}.) 
As a notational shorthand, I will write that implicit function
as $\SSS : \N \times \N \rightarrow N$. ($\SSS$ stands for ``self-simulation map''). As discussed above, $\SSS$ is computable.
In general, I will leave the second argument of $\SSS$ implicit, and just write the TM whose existence is
ensured by the self-simulation lemma as $\SSS(\Delta t)$. So the output of the self-simulating computer when
predicting the state of its own universe at time $\Delta t$ when the initial state of its environment is $w_0$ is
$g(\Delta t, w_0, \SSS(\Delta t))$.

\begin{example}
Suppose you have a laptop which has a memory that can be extended dynamically by an
arbitrary finite amount, an arbitrary number of times.
You've got some program $n_0$ that was already loaded into
the laptop at iteration $0$. The first thing you then do is load into the input of that program (an encoding of) the current state of
your environment, i.e., of the rest of the universe external to your laptop, which is $w_0$. This changes $n_0$ to
some new value, $n_1$.

After this you physically isolate isolate your laptop, from the rest of the universe, and start running it. 
The function $g(\Delta t, w_0, n_0)$ gives the joint state of your laptop and the universe external to it at the time
$\Delta t$ into the future.
Note that in particular you, the being who is running your laptop, is part of the environment of that laptop.
So your initial state is specified in the value $w_0$.

The self-simulation lemma says that there is some program $n_0$ that your laptop could have started with such that under the dynamics of
the universe, it is guaranteed to halt at some finite time $\widehat{\TT}((n_0, w_0)$. At that time that it halts,
its output would be the joint state of the universe external to your
laptop \textit{and of the laptop itself} at that time $\Delta t$. All other variables in the laptop other than this output have
been reinitialized, to the values they had before $n_0$ was loaded onto the laptop, e.g., to be all blanks.
\label{ex:1}
\end{example}

Note that for any $k \in \N$, there are an infinite number of indices $i \in \N$ such that $T^i = T^k$. This
means we can trivially modify the proof of \cref{lemma:2}
to establish that there are an infinite number of initial states of the computer, $n_0$, such that that computer simulates 
the full universe, including itself. In \cref{ex:1}, this is reflected in the fact that there are an infinite number of precise
programs all of which perform the came computation as the program $n_0$. Note though that in general, those different
programs will require different numbers of iterations to complete their computations of the future state of the universe
they are in.

Another important point is that that the self-simulation lemma holds for any $\Delta t$. 
Whatever time $\Delta t$ we pick to evolve our universe to, there is a program we can use to initialize the computer
subsystem so that it will calculate the joint state of the universe at that time. So in particular,
if $\TT(n_0, w_0) > \Delta t$, then the computer calculates its own state at that time $\Delta t$, a state
it had along the way while it was calculating what the joint state would be at that time. 

It's also worth pointing out
that the proof of the self-simulation lemma does not require the full power of the recursion theorem. That theorem applies to 
any function so long as it is partial computable. However, the self-simulation lemma only
needs to use it for the specific case of the evolution function, which is in fact a total computable function.

Note as well that the self-simulation lemma 
``hard-codes'' the time $\Delta t$ into the program $n_0$ that will run on the computer. 
If we change $\Delta t$, then in general we will also changes $n_0$. This is what allows us
to define the function $\SSS(\Delta t)$ (for implicitly fixed $g$).
Note though that $\Delta t$ does not exist in the universe as a physical variable, in addition to 
the variables $w$ and $n$. It is
merely a parameter of the evolution function that determines how we wish to initialize the (physical) variables of the computer.
So the cosmological universe does not perform that calculation of $\SSS(\Delta t)$ for some physically specified $\Delta t$,
and then use that value $\SSS(\Delta t)$ to initialize $N$. See~\cref{sec:why_time_not_in_V} for more discussion of this point.

\subsection{Difference between self-simulation lemma and previous work}

It's important to distinguish the self-simulation lemma from earlier concepts considered in the literature.
First, the traditional versions of the simulation
hypothesis discussed before did not involve the possibility that the beings running a 
simulating computer might be simulating their own universe, including themselves. Self-simulation appears to be novel
(\textit{pace} informal musings like the one quoted in this paper's epigraph).

Note as well that the self-simulation lemma is not a reformulation of the old warhorse, central to so much of computer science theory
and the foundations of math, of a mathematical object  ``referencing itself''. To clarify the difference,
suppose we feed a universal TM $U$ an encoding of itself, $\langle U \rangle$, along with a finite bit string
$w$ and then run that TM $U$. In other words, using $\langle a, b\rangle $ to indicate an encoding of a pair of bit strings, $a$, $b$, into
a single bit string, we are considering the case where $U$'s input tape initially 
contains the bit string $\langle \langle U\rangle, w\rangle $. In this situation $U$ is ``referencing itself''.
In particular, $U(\langle \langle U\rangle, w\rangle )$ is computing the output bit string that it itself would calculate if
its input tape were initialized with just $w$. Stated differently, the ``inner'' $U$, whose behavior if run on $w$ is being calculated, is the
same as the ``outer'' $U$ running with $\langle \langle U\rangle, w\rangle $ on its input, with this same $w$.

There are several major differences between such self-reference and the self-simulation lemma.
Most obviously,  there is no notion of simulating the laws of physics arising in the computation $U(\langle  \langle U\rangle, w\rangle )$.
In contrast to the both the self-simulation lemma and the traditional version of the simulation hypothesis, neither the inner
nor the outer $U$ is simulating the dynamics of our computational universe. 

Perhaps more importantly, $U(\langle  \langle U\rangle, w\rangle )$ does {not}
calculate what it itself would do \textit{if provided the string $\langle  \langle U\rangle, w\rangle $ as input}
(rather than be provided $w$ as its input). So it is not simulating its \textit{actual}
behavior, but rather some counterfactual behavior. Indeed, for $U$ to simulate its actual behavior on its actual input string, 
we would need to feed $U$ an infinitely long bit string, defined by an infinite regress: 
\eq{
\langle \langle U \rangle, \langle \langle U \rangle,  \langle \langle U \rangle, \ldots \rangle\rangle \rangle\rangle  \rangle\rangle )
}
(One can see this simply by iteratively expanding $w$ into $\langle \langle U \rangle, w\rangle )$.)
It would be physically impossible for any computer $\DD$ implementing $U$ to finish its computation 
 in finite (physical) time if it starts with this infinite string as its input. Moreover, there is no way to get information about the rest of the physical
universe \textit{outside} of that computer $\DD$ into the input to $\DD$ (since such information would need to be
appended at the end of the infinite string $\langle \langle U \rangle, \langle \langle U \rangle,  \langle \langle U \rangle, \ldots \rangle\rangle \rangle\rangle  \rangle\rangle)$). None of these problems apply with the self-simulation lemma.

It is also important to distinguish the self-simulation lemma from the age-old observation that if i) the 
universe extends infinitely spatially; ii) the physical constants (and more generally physical laws)
do not vary across that infinite space; iii) the initial conditions of the regions inside the backward local
light cones across the full universe are IID random; then somewhere else in this universe there is a copy of each of us, identical
down to a fine level of detail.\footnote{In fact, as has often been pointed out, we don't need to assume the IID random property.
If the universe is infinite then the randomness
of quantum mechanics means that there \textit{must} be such a copy of each of us --- indeed, there must be an infinite number of
such copies.}. This simple statistical phenomenon does not result in \textit{exact} identity between any of
those copies of us and us. More importantly, perhaps the most striking aspect of the self-simulation lemma
is that if it applies to us, it would mean that \textit{we} are running a program on a computer that is simulating ourselves, as we
run that simulating program. There is as direct coupling between ourselves and our indistinguishable copy as possible; one of
us is directly controlling the other one of us, but those two versions of us are one and the same. Phrased differently, one of
of us is the parent, in a deep and fundamental sense, of the other. In fact, we are both the parent and the child.
Indeed,  if the self-simulation lemma holds, then
there is an infinite regress of copies of you, residing inside the successive layers of nested dolls of computers
simulating computers. No such nested set of copies of an individual arises under the ``age-old observation''.

Another function considered in the literature that in some respects
resembles the function $g$ of the self-simulation lemma
is the Von Neumann constructor. That is a configuration of neighboring states in a 
particular cellular automaton that is capable of constructing an
identical copy of itself as the cellular automaton evolves. However, a Von Neumann constructor releases the copy of itself once it has created
it, and that copy then has a completely independent existence, undergoing different dynamics. In fact, there is
no time at which the Von Neumann constructor and that of the copy of itself that it constructs
undergo identical dynamics. The Von Neumann constructor does not ``run'' the copy of itself that it makes,
in the sense of the function $g$ in the self-simulation lemma.

\section{Time delay, cheating computers, and self-simulation}
\label{sec:time-delay}

In this section I present a few of the particularly elementary mathematical properties of self-simulation.

\subsection{Necessity of time delay in self-simulation}

One might suspect that there must be a delay between the time $\Delta t$ into the future that a self-simulating computer
is simulating, and the time at which it completes that simulation. Intuitively, the idea would be that,
assume instead that $\TT( \Delta t, w_0, \SSS(\Delta t)) = \Delta t$. This would imply that
\eq{
n_{\Delta t} &= (\langle w_{\Delta t}, n_{\Delta t}\rangle) \\
	&= (\langle w_{\Delta t}, \langle w_{\Delta t}, n_{\Delta t}\rangle)  \\
	&= (\langle w_{\Delta t}, \langle w_{\Delta t}, \langle w_{\Delta t}, n_{\Delta t}\rangle \rangle \rangle)  \\
	& ...
}
By the pigeonhole principle, the lengths of the encoded strings in this sequence must grow infinitely long.
Therefore $n_{\Delta t}$ would have to be an infinitely long string. But no UTM can write an infinite number of bits
onto its output tape in a finite number of iterations. This implies that the equality cannot be satisfied.

One could make a counter-argument though. 
One response to this intuitive argument is to point out that our cosmological universe in many ways evolves more like a parallel 
computer rather than a serial one, so that for example the state space of $N$ could evolve like an infinite one-dimensional Cellular automaton.
That would see to allow an infinite number of operations to occur simultaneously --- disproving the intuitive
argument above. The computational implications of there being parallel rather than serial dynamics in our universe are subtle though --- see \cref{app:CAs_not_TMs}. 

Even if we do restrict attention to a serial computer though, suppose
we encode the bit-string $(w_{\Delta t}, \langle w_{\Delta t}, \langle w_{\Delta t}, n_{\Delta t}\rangle \rangle ), \ldots)$ as the
output generated by a simple computer
program, a program which would require very few bits to write down. So while 
$(w_{\Delta t}, \langle w_{\Delta t}, \langle w_{\Delta t}, n_{\Delta t}\rangle \rangle ), \ldots)$ is infinitely long, it can be encoded as a finite
computer program (a short program, in fact). The simulation computer \textit{could} write such a finite program onto its output
tape in a finite time. On the other hand though, there is no computable function that
can decode that version of the infinite string $(w_{\Delta t}, \langle w_{\Delta t}, \langle w_{\Delta t}, n_{\Delta t}\rangle \rangle ), \ldots)$
which has been encoded as a computer program. Any program that tries to do this would never halt.

These arguments and counter-arguments are resolved in the following lemma:
\begin{lemma}
For all cases where a computer $V$ is simulating itself,  and for all associated $\Delta t, w_0$, 
\eq{
\TT( \Delta t, w_0, \SSS(\Delta t)) > \Delta t  \nonumber
}
assuming $|W| > 1$. 
\label{lemma:min_delay}
\end{lemma}
\begin{proof}
Hypothesize that there is some $\Delta t, w_0$ such that
\eq{
\TT(\Delta t, w_0, \SSS(\Delta t)) \le \Delta t  
\label{eq:26}
}
Then by \cref{def:sim}, for all $t \ge \TT(\Delta t, w_0, \SSS(\Delta t))$,
\eq{
g(t, \WW(\Delta t, w_0, \SSS(\Delta t)), \NN(\Delta t, w_0,  \SSS(\Delta t))_{\underline{N}} &=
	g(t, \WW(\Delta t, w_0, \SSS(\Delta t)), \SSS(\Delta t))_{\underline{N}}  
}
So in particular, we would have
\eq{
	g(\Delta t, \WW(\Delta t, w_0, \SSS(\Delta t)), \SSS(\Delta t))_{\underline{N}}  &= \langle g(\Delta t, w_0,  \SSS(\Delta t)) \rangle  
\label{eq:28}
}

Using the fact that we're doing \textit{self}-simulation, $\SSS(\Delta t) = n_{\Delta t}$, and
$\WW(\Delta t, w_0, \SSS(\Delta t)) = w_0$. Plugging this into \cref{eq:28},
\eq{
	g(\Delta t,w_0, n_{\Delta t})_{\underline{N}}  &= \langle g(\Delta t, w_0, n_{\Delta t}) \rangle  
}
However, again using the fact that we're doing self-simulation,
$g(\Delta t, w_0, n_{\Delta t})_{\underline{N}}$ just equals $n_{\Delta t}$. Combining,
\eq{
n_{\Delta t}  &= \langle g(\Delta t, w_0, n_{\Delta t}) \rangle  \\
	&= \anglebracket{w_{\Delta t}, n_{\Delta t}}
\label{eq:32}
}

Next, recall that in this paper I assume the encoding $\anglebracket{., .}$ produces strings whose lengths are non-decreasing functions 
of the lengths of its two arguments. (See~\cref{sec:notation}.)
So if $W$ has more than one state (and therefore $w_{\Delta t}$ is at least a bit long), 
$|\anglebracket{w_{\Delta t}, n_{\Delta t}}| > |n_{\Delta t}|$.
In this case \cref{eq:32} is a contradiction. So our hypothesis must be wrong, which establishes
the lemma for the case $|W| > 1$.
\end{proof}

\cref{lemma:min_delay} means that
for all $w_0$, $\TT(\SSS(\Delta t)), w_0, \Delta t)$ has no upper bound as $\Delta t$ grows, and must always exceed $\Delta t$. It
does not means that $\TT(n^*, w, \Delta t)$ is an everywhere increasing function of $\Delta t$ though --- 
$\TT(\SSS(\Delta t)), w_0, \Delta t) + 1$ can
be less than $\TT(\SSS(\Delta t)), w_0, \Delta t)$, so long as it's greater than $\Delta t$. 

One could of course forbid this possibility, simply by requiring that  $\TT(\SSS(\Delta t)), w_0, \Delta t)$ is not decreasing
as a function of $\Delta t$. There are other ways to address this issue as well. One, mentioned in \cref{sec:framework}, is
to modify \cref{def:sim} so that the simulating computer does not just produce a single future state of the
universe being simulated, but rather produces an entire (finite) sequence of future states of the simulated universe, in
order. A related way to address this issue is addressed next, in \cref{sec:restrictions}.

\subsection{Restrictions to impose on the evolution function}
\label{sec:restrictions}

There are several additional restrictions it will sometimes be natural to place on the evolution function. It would not
change the main results presented below in~\cref{sec:sim-lemma,sec:self-sim-lemma} to impose those restrictions.
However, it is helpful to invoke them in certain parts of the subsequent analysis of those results. 

We begin with the following definition:
\begin{definition}
A universe $V$ has a \textbf{(stationary) Markovian} evolution function $g$ if for all $\Delta  t > 0, w\in W, n \in N$, 
\eq{
g(\Delta t, w_0, n_0) = \gamma^{\Delta t}(w_0, n_0) \nonumber
}
for some function $\gamma : W \times N \rightarrow W \times N$
\end{definition}
\noindent The dynamics of a stationary Markovian universe can be expressed as a time-translation invariant function $\gamma$
repeatedly running on its own output, i.e., as an iterated function system. 
In practice, we are often interested in universes whose dynamics is {Markovian}. In particular, we humans  believe
that our actual cosmological universe has this property.

Another restriction is especially natural to impose when considering universes that simulate
themselves. This is the restriction that the computer in that universe and its environment do
not interact after some iteration $k > 0$. Arguably, without this restriction, we have no assurances that
the computer is \textit{simulating} the future state of its environment from times $k$ to $\Delta t$, rather
than just observing the state of its environment at time $\Delta t$. We can capture this restriction in the following definition:

\begin{definition}
\label{def:shield}
Fix $\Delta t$, and choose some integer $k$ such that $0 < k < \Delta t$.
A universe $V$ with a Markovian evolution function is \textbf{shielded (after iteration $k$)} if
for all $w, w_0, n_0$,
\eq{
[\gamma^{\Delta t - k}(w, n_k)]_N
}
is independent of $w$, where we define
\eq{
n_k := [\gamma^k(w_0, n_0)]_N
}
\end{definition}

\subsection{Cheating self-simulation and how to prevent it}
\label{sec:cheating_computers}

{There are several ways that the self-simulation lemma can be met that can
be viewed as ``cheating''. Perhaps the most egregious is the scenario in the following example:}

\begin{example}
\label{ex:3}
Suppose the computer's initial state $n_0$ is blank, and does not evolve until time $\Delta t$. Then
at time $\Delta t$, the value $w_{\Delta t}$ is copied onto the state of the computer (e.g., by the computer observing
that state of the environment at that time, or some beings in the external universe
overwriting the state of the computer at that time). This results in the computer state $n_{\Delta + 1} = w_{\Delta t}$, which was
precisely the state of $v_{\Delta t}$.
\end{example}

The scenario described in \cref{ex:3} can be prevented by requiring that the computer is shielded for all times
after some $k \ll \Delta t$. However, even if we require shielding, together with the  associated requirement
of Markovian evolution, the self-simulation lemma can still be satisfied if the computer essentially does nothing up to the time $\Delta t$,
so that its state at that time does not depend on $w_0$. This is illustrated in the following example.\footnote{As was mentioned in
\cref{sec:framework}, in this subsection
I need to explicitly write $n_t = (\underline{n}_t, r_t)$, rather than use the shorthand of identifying the state of the computer
$n_t$ with the state of its tape, $\underline{n}_t$ which is used in most of the rest of this paper.}

\begin{example}
\label{ex:2}
For simplicity, I describe this scenario as though the computer is a single-tape UTM.
Suppose that the initial state $\underline{n}_0$ of the tape of the TM is $\Delta t$.
The first thing that happens is the tape is provided the initial state of the computer's environment, $w_0$ (implicitly followed
by an infinite string of blanks). Suppose
that in addition, there is a ``counter'' variable $c$ that is initialized with the value $0$ that is appended to $w_0$ on the
tape. So the initial state of the tape is $\underline{n}_0 = (\Delta t, w_0,  0)$, the initial state of the full TM is
${n}_0 = [\Delta t, w_0,  0; r_0]$, and the initial state of the full universe is $v_0 = ([\Delta t, w_0,  0; r_0], w_0)$.

The TM evolves shielded from its environment from now on. In all 
subsequent iterations up to $\Delta t$, $\underline{n}_t$ does not change, while $c$ increments by $1$ in each of those iterations.
Then when the counter reaches $\Delta t$, it stops incrementing. At this time the contents of the tape of the TM
is $(\Delta t, w_0,  {\Delta t})$, with the other variables of the TM (its state and its head's position) having some value
$r_{\Delta t}$. So the entire ID of the TM at this moment is  $[ {\Delta t}, w_0,  {\Delta t}; r_{\Delta t}]$, and therefore the state of
the full universe is $([\Delta t, w_0,  {\Delta t}; r_{\Delta t}], w_{\Delta t})$.

Next the computer makes a second copy of $w_0$ and appends that together with $r_{\Delta t}$
to the end of its tape, so that it now has $(\Delta t, w_0,  {\Delta t}, r_{\Delta t}, w_0)$ on its 
tape.\footnote{Recall the discussion in \cref{sec:framework} of the fact that it may prove convenient to augment our TMs with
a special instruction that copies the contents of an arbitrarily large portion of $V$ to another portion of $V$ in a single
iteration; that extra instruction could be used here, for calculational convenience, so that we don't have to account for
the number of iterations it would require a non-augmented TM to copy over that entire input, working on only a single
variable in the multi-dimensional state space $W$ at a time.}
It then computes $w_{\Delta t}$ from that copy of $w_0$, so that at some later time
$t_2 > \Delta t$ its tape is $(\Delta t, w_0,  {\Delta t}, r_{\Delta t}, w_{\Delta t})$. 

At this time the state of the full universe is $v_{t_2} = ([\Delta t, w_0,  {\Delta t}, r_{\Delta t}, w_{\Delta t}; r_{t_2}], w_{t_2})$.
Note though that the value of the tape of the TM at $t_2$ is identical to the state of the entire universe at $\Delta t$.
So $n_{t_2} = g({\Delta t}, w_0, n_0)$. We therefore satisfy the self-simulation lemma by choosing $\TT(\Delta t, w_0, n_0)
= t_2$.
\end{example}

One might argue that in a certain sense the algorithm used in \cref{ex:2} for self-simulation is a milder
type of ``cheating'', since the computer only increments
a counter up to time $\Delta t$, not trying to evolve $w_0$ at all. However, we can nest that algorithm within itself
an arbitrary number of times, so that $w_0$ \textit{is} being evolved continually, not just after reaching the time $\Delta t$.
This is illustrated in the following extension of \cref{ex:2}, which shows how a shielded computer can produce a subsequence
of an entire trajectory of states of the full universe, computing those states one after the other, in the proper time order:

\begin{example}
\label{ex:4}
As in \cref{ex:2}, treat the computer as a single-tape TM. However, do not have any value $\Delta t$ on the
tape of the TM at $t_0$. In addition, there is no counter variable. Other than that, we run the 
exact same algorithm as in \cref{ex:2}, as though $\Delta t$ had been set to $1$, and there was no need for
counter incrementing. So the first thing that happens is the state of the tape gets overwritten with the value $w_0$.
Suppose that this copy operation completed at some iteration $t_1 > 0$. So the full ID of the TM at $t_1$ is $[w_0; r_{t_1}]$,
and the state of the universe then is $([w_0; r_{t_1}], w_{t_1})$. 

Next the TM appends $r_{t_1}$ to the end of its tape, and then 
copies $w_0$ to after that. When this is done it computes $w_{t_1}$ from the copy of $w_0$,
overwriting that copy. Supposing it completes this 
at $t_2 > t_1$, the state of its tape at $t_2$ is $(w_0, r_{t_1}, w_{t_1})$, the full ID of the TM then is $[(w_0, r_{t_1}, w_{t_1}); r_{t_2}]$, 
and the state of the universe then is $v_{t_2} = [(w_0, r_{t_1}, w_{t_1}); r_{t_2}], w_{t_2})$. So the state of the TM at $t_2$
is the state that the entire universe had at $t_1 < t_2$.

At this point the TM appends the value $r_{t_2}$ to its tape,
and then appends a copy of the state $w_{t_1}$ that was stored on its tape to the end of its tape.
It then uses that copy to compute $w_{t_2}$. Assuming it completes that computation at iteration $t_3 > t_2$, at iteration
$t_3$ the state of the tape is $(w_0, r_{t_1}, w_{t_1}, r_{t_2}, w_{t_2})$, the ID of the TM is 
$[(w_0, r_{t_1}, w_{t_1}, r_{t_2}, w_{t_2}); r_{t_3}]$, and the
state of the full universe is $([(w_0, r_{t_1}, w_{t_1}, r_{t_2}, w_{t_2}); r_{t_3}], w_{t_23})$. 
In particular, the state of the tape at $t_3$
is the state of the full universe at $t_2 < t_3$.

The computer keeps repeating this process, without ever halting. (Or alternatively, it can halt after some arbitrary, pre-fixed
number of iterations, using a counter variable to count iterations that is stored in $R$.)
As it does so it computes the full states of the universe at the iterations $1, t_1, t_2, t_3, \ldots$, outputting those computations
at the iterations $t_1, t_2, t_3,\ldots$, respectively, where $t_1 < t_2 < t_3 < \ldots$.

\end{example}

I refer to the procedure run in \cref{ex:4} as \textbf{(greedy) nested} simulation of a trajectory of states, with
the sequence $\{t_1, t_2, \ldots\}$ called the \textbf{simulation time sequence}.\footnote{The qualifier
``greedy'' indicates the fact that each successive computation is written to the tape of the TM as early as possible.
Technically, greedy nested simulation means that for all $t_i$, $t_{i+1} - t_i$ is as small as possible.} In this case the
nested simulation is applied
by a universe to itself, a special case I refer to as \textbf{greedy nested self-simulation}.
Note that in nested self-simulation, while the number of iterations to compute the state of the universe at a time $t$ is an increasing
function of $t$, it is a partial function. Only a sub-sequence of the full trajectory of states of the universe
defined by the simulation time sequence, $\{v_{t_1}, v_{t_2}, v_{t_3}, \ldots\}$, is computed. 

The nested self-simulation only places a sequence of pairs $(r_{t_i}, w_{t_i})$ onto the tape, 
never separating the two elements of such a pair in its output. In addition, recall from above we can almost 
always treat $N$ as synonymous with $\underline{N}$, with \cref{ex:4}
being the only instance in this paper in which we explicitly distinguish the $\underline{N}$ and $R$ components of
$N$. Given all this, one might think that
formally, we could absorb the variable $R$ into the variable $W$, leaving only $\underline{N}$ in
the computer variable $N$. Doing that would change $W$ into the environment of $\underline{N}$, 
not of $N = \underline{N} \times R$. 

This would simplify the notation of this paper. However, \cref{def:shield} requires
that $R$ and $\underline{N}$ both lie in $N$. If we absorbed $R$ into $W$, then in nested self-simulation the computer
$N$ (which would now only consist of the tape $\underline{N}$) would have to interact with its external environment for all iterations,
never evolving autonomously. So we could not require that the computer be shielded from its environment.

%
%

\section{Philosophical issues raised by the simulation and self-simulation lemmas}
\label{sec:philo_implications}

\subsection{Who am I?}

The simulation and self-simulation lemmas have some interesting philosophical aspects. Most obviously, suppose that
the PCT and RPCT both apply in our particular computational universe.
Then not only might we be a simulation in a computer run by aliens in a universe
that our universe supervenes on  --- we and our entire universe might be
a simulation in a computer \textit{in our very universe}. (This is not an issue considered in the earlier
literature on the simulation hypothesis.) It might be that we comprise a portion of the universe external
to the simulation computer, i.e., our state \textit{in toto} at time $t$ is specified by $w_t$, and our dynamics
is exactly given by $g$, and therefore we and our dynamics would {also} be exactly given
by the dynamics of $n$. In other words, we would be both in the universe external to the computer,
and in the simulation being run on the computer. And importantly, \textit{there would be no possible
experimental test we could perform} that could distinguish ``which of those two entities we are''. Our
existence would be duplicated; we would \textit{be} a duality, in all respects.

In that particular scenario, we are not the ones running the simulating computer.
However, if we ever in the future gain the capability of building and running simulation computers, 
then by the self-simulation lemma, at that time we might even be simulations in a computer
that we ourselves run! In such a situation, since the part of our universe containing us, $W$, is being reproduced in exact 
detail inside the computer that is simulating the dynamics of $W$, the version of us inside 
the computer is itself running a computer, that in turn is simulating us running a computer in exact detail.

In other words, we the humans running that physical computer inside of $W$,  might ``be'' either
those people running that physical computer --- or we might be the people inside the physical computer who are evolving
as the simulation runs, and who are indistinguishable from the ``other'' humans who are running that physical computer.
By the RPCT, there is no conceivable physical test, no observable value, that could tell us which of those
two dynamic processes ``is'' us. So in a non-Leibnizian, empirically meaningful sense of the 
term, we could ``be'' either one of those two evolving objects. We could even both, and would never know. 

This conundrum raised by the self-simulation lemma concerning the concept of ``identity''
is in some senses reminiscent of the ``the boat of Theseus'' concept. Going beyond that concept
though, the self-simulation lemma considers a situation where one object that is directly controlling the other, as both evolve. 
By construction we cannot distinguish between
the possibilities that we are the controller object or we are the controlled, as both of them evolve. 
There is no such splitting of identity among two simultaneously evolving objects in the boat of Theseus scenario. 

As a related point, loosely speaking, one can define ``conscious experience'' of a person
as their thinking about their own thinking. If we adopt that definition, and modify the RPCT thesis appropriately, then
 we could use the associated modified version of the self-simulation
lemma  to establish the formal possibility of
conscious experience. Rather than apply the original version of the
lemma to a physical computer's simulating itself as it simulates itself, we would apply this modified version ot the lemma
a physical brain that performs computations (``thinks'') about those computations (``thoughts'') it is performing.

\subsection{Running (self-)simulation using fully homomorphic encryption}

Another interesting set of issues arises if there is one universe $V = W \times N$ that simulates a second universe
$V' = W' \times N'$, but that simulation is a fully homomorphic encrypted (FHE) version 
of the evolution function of 
that second universe.\footnote{Recall that in FHE you have an algorithm that runs on some encrypted data, producing
a result that when decrypted is identical to the result you would get if ran an associated algorithm on the original, pre-encrypted data, without
any encryption of any sort. So in order to use FHE encryption to run a program in an encrypted fashion,
all you need to do is have the encrypted data specify that program, and have the algorithm running on the encrypted
data be a UTM.} In other words, the program $n_0 \in N$ could be an FHE version
of the program simulating $V'$ that the beings who are running that computer $N$ want it to compute.
So those beings would need to use a decryption key to understand the result of their computer's simulation of the evolution of $V'$.

In this case, as a practical matter, if the simulator beings 
lost the decryption key, then they would not be able to read out the 
results of their simulation. This would be the case even though that simulation would in fact be perfectly
accurate. Or to be more precise, those simulator beings \textit{could} read the
results of their simulation --- but would require their expending a huge amount of computational resources
to do so.\footnote{As an aside, 
recall the common supposition that the sequence of events in our universe
must have low Kolmogorov complexity, in order for it to contain a pattern that is evident to us, 
so that it ``counts as having been generated by mathematical laws, rather than 
just being a lawless, random sequence''. Note though that
running  a program via FHE rather than running it directly does not change its Kolmogorov complexity. (Though
to run it via an FHE \textit{and then also decrypt the results} would result in a composite program with slightly larger Kolmogorov complexity.)
So we could have a sequence of events that
 \textit{appear} to be purely random, to us (and so would not ``count as mathematical laws''), even though they have
low Kolmogorov complexity. In such a situation they have low complexity,
but \textit{we} cannot distinguish them from a sequence of events with
high Kolmogorov complexity. One might argue that even if low Kolmogorov complexity of the sequence is not a sufficient condition
for it to be considered lawful, it is still a \textit{necessary} condition
for it be considered lawful. However, Chaitin's incompleteness theorem says it is impossible
to prove that any sequence with Kolmogorov complexity above a very small value actually has that Kolmogorov complexity.
So we can never prove that such a necessary condition is violated.}

On the other hand, since $V'$ obeys the PCT there is no sense in which the beings being simulated
could know that they are being produced in a simulation. Life would appear ``normal'' to them, with no
``randomness'' of any sort. So in particular,
there is no way that they would be able to distinguish between being produced in a simulation being made via an FHE algorithm, 
or instead in some simulation that is easier to understand. As an example, suppose we are a simulation, so that the laws of physics 
we perceive are simply
the evolution function $g'$ of our universe. In this situation, we would not be able to distinguish the case where we are being run
on a simulation program that directly implements the laws of physics, in the straight-forward way (cf.~\cref{sec:example_V_our_universe}), 
or are instead being run on a FHE version of the laws of physics. Moreover, in the latter case, if we could actually
somehow see our universe's evolution from the perspective of the beings who are running
the program producing us, we would not be able
to distinguish the laws of physics controlling our universe in that simulation from completely random noise. In
this sense, the actual laws of physics in our universe might in fact be pure noise --- and we would not be able to
tell the difference.

Similar consequences arise if we consider self-simulation. We might be simulating ourselves, but have done
so with an FHE version of ourselves. Similarl to the possibility discussed above, suppose this is the case, but that we 
have misplaced the decryption key. In this scenario, we are a simulation
that we ourselves are running --- but we cannot understand that simulation of ourselves, a simulation which \textit{is}
us.

\section{The simulation graph}
\label{sec:simulation_graph}

\cref{sec:philo_implications} contains a brief discussion of the philosophical issues that arise if we
consider simulation involving more than two universes.  There are also interesting mathematical aspects
to such a situation. In particular, the graphical structure of universes simulating universes can be quite interesting.

I start in this section with some preliminary remarks concerning that graph. Then in \cref{sec:math_implications}
I discuss some other open mathematical questions.

\subsection{The graph of simulations and self-simulations}

I begin with the simple observation that simulation is a transitive relation:
\begin{lemma}
If $V$ simulates $V'$, and $V'$ simulates $V''$, then $V$ simulates $V''$.
\label{lemma:transitivity}
\end{lemma}
\begin{proof}
The proof parallels that of \cref{lemma:sim}.
By hypothesis, the evolution function $g''$ of $V''$ is computable,
and there exist associated functions ${\TT}_{V',V''}$, ${\WW}_{V',V''}$, ${\NN}_{V',V''}$
that obey the RPCT properties for $V'$ for (a set $K'$ that includes the number $k'$ coding for) the TM $T^{k'}$ that
implements the evolution function $g''$, where the argument $y'$ of $T^{k'}$ is set to $\anglebracket{\Delta t'', w''_0, n''_0}$
for any $\Delta t'' \in \N$.

Similarly, by hypothesis there exist functions ${\TT}_{V,V'}, {\WW}_{V,V'}, {\NN}_{V,V'}$
that obey the RPCT properties for $V$ for (a set $K$ that includes the number $k$ coding for) the TM $T^k$ that
implements the evolution function $g'$, where the argument $y$ of $T^k$ is set to $\anglebracket{\Delta t', w'_0, n'_0}$
for any $\Delta t'' \in \N$. So in particular, $V$ has this property where the argument 
$y$ of $T^k$ is set to 
\eq{
\anglebracket{{\TT}_{V',V''}(\Delta t''), {\WW}_{V',V''}(w''_0), {\NN}_{V',V''}(n''_0)}  \nonumber
}
\end{proof}

A \textbf{(bare) simulation graph} $\Gamma$ is defined as a directed graph whose nodes are universes
where there is an edge from $V$ to $V'$ iff $V$ simulates $V'$. In light of the transitivity
of simulation, for any node $V$ in a simulation graph $\Gamma$, and any node $V'$ on a directed path leading
out of $V$, $\Gamma$ must contain an edge from $V$ to $V'$. In general though, if $V'$ and $V''$ are two universes
that are both descendants of $V$ in the graph $\Gamma$, it need not be the case that one of them can simulate the other.
In addition, due to the possibility of self-simulation, the simulation graph
need not be a simple graph; it might contain edges that point to the same
node that they came from. 

Indeed, suppose we restrict attention to such a simulation graph universes such that there is a directed
edge from any node $V$ in the graph to any other node $V' \ne V$.  Then the simulation relation cannot be
a partially ordered set. Given this, suppose that all of the nodes (universes) have a finite $W$ and an evolution
function that obeys the PCT. As pointed out in \cref{sec:PCT}, the set of such universes is countably infinite,
and so we cannot assign a uniform probability distribution to that set. However, since the elements of the set
are not partially ordered in the simulation graph, we also cannot assign a Cantor measure over the elements
of the set, if we wish to use the simulation relation to fix how to assign such a measure to the nodes in
the graph. This establishes the claim made in the introduction, that it is not possible to use a 
Cantor measure to assign probabilities to a very naturally defined set of universes (at least, it's not
possible if we try to use the simulation relation to fix the measure).

In general, a bare simulation graph could contain computational universes that exist in the same 
cosmological universe (see \cref{sec:framework}), obeying the same laws of physics, the same initial
conditions, etc. It might also contain computational universes that exist in different cosmological
universes. These two cases can be intermingled as well. 

Note that in general there might also be multiple 
edges coming \textit{in} to each node. If we are such a node, that would mean that that we could
simultaneously be the simulations being run by more than one set of beings in other cosmological universes. 
Another possibility is for there to be multiple beings in a single cosmological universe all of whom are running a simulation that is us. 

It's worth briefly commenting on how this second possibility might come about in our particular cosmological universe.
One way is if those beings are all running the simulations that are us at the same (cosmological, co-moving) 
time. In such a situation
we would have ``split'' identities, but at least they would all exist at the same moment in time in our universe. 
Alternatively though, those beings could be running simulations that are us, but are doing so at different times (i.e., 
where there is no co-moving frame that contains all of those beings at the moments they are running the
simulations). In such a situation,
our separate (but identical) selves would all exist at a different moment of time. (Though of course, those versions of us could not
have seen that difference in times, at least not yet, since after all, they are \textit{identical}.) 

The simplest way either of these situations could be arranged is if each of the beings running simulations of us exists
in a region of the same cosmological universe, where each region
is causally disconnected from the regions containing the other beings, and whose backward
light cone doesn't intersect the backward light cones of the other beings simulating us. These restrictions
would prevent any complications
from the need to have the intersections of the light cones not prevent those beings from all simulating
us. (This is true of both the case where all the beings simulating us exist at the same time, and the case where
some of them exist at different times.)


In addition to being simulated by multiple other sets of aliens, we could ourselves
be running simulations that are us, while perhaps simulating some of those beings
who are running simulations of us. In this structure there would be edges that are loops from us into us,
and edges from those other universes into us, perhaps together with some edges from us into
those other universes.

Suppose we restrict the simulation graph $\Gamma$ to only contain universes that can simulate themselves,
Suppose as well that we restrict the graph so that there is at most a single edge from any universe to $V$ to $V'$ (
where $V'$ may or may not differ from $V$).
Then the directed edges in $\Gamma$ form a reflexive, transitive relation, i.e., they form a preorder. In general, the relation 
provided by the edges in that simulation graph can include both pairs of nodes that are symmetric under the relation
and pairs of nodes that are anti-symmetric under the relation. So while it is a preorder, those edges 
need not provide either an equivalence relation or a partially ordered set. 

There exist many kinds of equivalence classes that could apply to a simulation graph, depending on
what universes it contained. Most obviously, we could always divide those universes into 
equivalence classes where all computers in a class can simulate one another. If in addition
the edges form a linear order, then all computers in a class can also simulate all computers in a 
class that is lower (according to the $\le$ ordering). But no universe can contain a computer that simulates a universe in
a higher equivalence class.


\subsection{Time-ordered and time-bounded simulation graphs}

If we consider a set of computable universes all of which obey the RPCT for $K = \N$, then the
simulation graph is trivial: it is a fully connected graph. 
However, even if we're considering a set of computable universes all of which obey RPCT for $K = \N$, there might
still be nontrivial structure in the graph given by placing an edge from $V$ to $V'$ only if $V$ simulates $V'$ 
sufficiently quickly. More formally, we can consider the \textbf{time-bounded simulation graph} in
which there is an edge from $V$ to $V'$ iff $V$ simulates $V'$
{with a function $\TT(\Delta t', w', n')$} that obeys some bound in the worst-case over all pairs $(w', n')$
in how fast it can grow as a function of $\Delta t'$. For example, one could consider the variant of a simulation
graph given by only placing an edge from $V$ to $V'$ if for all pairs $(w', n')$, $\TT(\Delta t', w', n')$ grows 
at most polynomially with $\Delta t'$. Other kinds of nontrivial simulation graphs arise if we consider 
the scaling of the resources (time, memory, etc.) needed to compute the functions $\TT, \WW, \NN$, or
in the case of self-simulation, $\SSS$. (See also the discussion in \cref{sec:discussion} of time-minimal
simulation functions, their scaling properties, and computational complexity theory.)

Next, consider any two nodes $V$, $V'$
connected by a directed path in a (bare) simulation graph. $\TT_V(\Delta t', w', n';)$ will differ from $\Delta t$. 
In this sense, $\TT(\Delta t, w, n)$ provides a well-defined measure of ``the speed of time
of the dynamics of a universe'', a speed of time that varies among the different universes along any path descended from $V$. 
Note though that this speed of
time is only defined for the universe being simulated, measured against time intervals in the universe doing the simulating.
Moreover, the speed of time might differ depending on $\Delta t$, i.e., $\TT(\Delta t, w, n) / \Delta t$ might vary depending
on $\Delta t$, even for a fixed $w, n$.

In the case of self-simulation, we know from \cref{lemma:min_delay} that in fact the speed of time is always
sped up in a universe being simulated by itself. 
Accordingly, I define a \textbf{time-ordered simulation graph} as any bare simulation graph where all edges $V \rightarrow V'$ are removed
where for at least one triple $(\Delta t', w', n')$, $\TT(\Delta t', w', n') < \Delta t$. (Whether or not $V = V'$,
as in self-simulation.) We do not allow edges in which the speed
of time slows down in time-ordered simulation graphs (though we allow time to speed up).

Suppose that in fact $\TT(\Delta t', w', n') < \Delta t$
for all $n', w'$ in all universes on the nodes of the graph. Then the time-ordered simulation graph cannot be cyclic. However, it could still ``spiral'',
in the sense that going along a directed path starting from a node $v_1$ could land on a node $v_N$ that
is an indistinguishable copy of (the universe evolving in) node $v_1$, except that the speed of time in $v_N$ is greater than that
in $v_1$. 
%


Recall as well from \cref{sec:features_self_sim} that when the conditions in \cref{lemma:2} hold for a universe $V$.
there are an infinite number of initial states of the computer, $n_0$, such that that computer simulates 
the full universe $V$, including itself.
This may have some interesting philosophical and mathematical implications. For example, it
suggests the possibility for me to run a computer which is simulating my entire universe for a given time into the future
$\Delta t$, for my actual environment $w_0$, and for a program $n_0$ that is computationally equivalent to
the actual simulation program I am using --- but where that program $n_0$
actually differs from the precise program I am using. (So it is not a perfect self-simulation, in that sense.)
This suggests replacing any single loop in the simulation graph(i.e., any single edge from a universe-node into itself)
with a set of multiple such loops, distinguished by the fact that they use different (but computationally equivalent)
programs.

\subsection{Weak RPCT}

In general we do not need to assume the full strength of the RPCT to prove a particular instance of 
either the simulation lemma or the self-simulation lemma.
In the case of the former, we just need to assume that the simulating universe can implement the evolution function of the universe being
simulated, i.e., can implement the particular TM specified by that evolution function of the universe being simulated. It is not necessary that it can
implement the Turing machine of any universe there is.
In the case of the latter, we just need to assume that the universe can implement the (Turing machine specifying the) 
computable function $\SSS(\Delta t)$. We do not need it to be computationally universal.

Accordingly, I say that the \textbf{weak RPCT (for a set $K \subset \N$)} holds for universe $V$
if \cref{def:RPCT} holds for $V$ after one replaces the requirement that the
functions $\widehat{\TT}, \widehat{\WW}$ and $\widehat{\NN}$ have the RPCT properties for \textit{all}
$k$, instead only requiring they they have the RPCT properties for all $k \in K$ (and evolution function $g$ of $V$).

The associated \textbf{weak simulation lemma} says that $V$ can simulate $V' = W' \times N'$ if $V'$ obeys the PCT and
$V$ obeys the weak RPCT for a set $K$ that includes three functions
$\widehat{\TT}, \widehat{\WW}$ and $\widehat{\NN}$ that have the RPCT properties for
all $(\Delta t', w' \in W', n' \in N')$ (for the associated evolution function $g'$).
With obvious generalizations, we can weaken the RPCT further, by restricting the set of $w' \in W'$
and / or the set of $\Delta t$ --- which corresponds to restrictions on the set of $y$ in \cref{def:RPCT}.
Transitivity of simulation 
(\cref{lemma:transitivity}) would still apply for a set of universes related this way. So the simulation graph for such
a set of universes would again be a preorder. 

Similarly, suppose that a universe obeys the weak RPCT for a set $K$,
where the solution $\SSS(\Delta t)$ for $V$ to simulate itself a time $\Delta t$ into the future
lies in $K$. In this case the universe does not obey the full RPCT, but it still is able to simulate
itself for that future time $\Delta t$. Accordingly, I call this the \textbf{weak self-simulation lemma}
for universe $V$. As with the simulation lemma, we can further weaken the RPCT so that we
limit the set of $w \in W$ and / or $\Delta t$ for which $V$ simulates itself.

One could extend the simulation graph by changing the definitions of the edges
to include these weakened versions of the simulation and / or self-simulation lemmas.
In particular, it might be of interest to investigate how the structure of the graph progressively changes 
as we progressively weaken those lemmas.

In a similar way, we could weaken the PCT, either instead of weakening the RPCT or in addition to
weakening the RPCT. This would result in yet another pair of lemmas, and another extension of the
simulation graph.

\subsection{Refinements of simulation graphs}

There are many other variants of simulation graphs that might be interesting to explore.
As an example, consider some computational universe  $V = W \times N$ where $W = \bigtimes_{i \in I} A_i$ for some set of spaces $\{A_i\}$.
Suppose that our computational universe is $V' = W' \times N'$, where $W' = \bigtimes {j \in J} A_j$ for some subset $J \subset I$.
Now suppose that the computer $N$ is running a simulation of us. (For example, there might some species of super-aliens
who are a part of the environment $W$ distinct from us, who are running the computer $N$ that way.) 
In this case there would be two instances of our universe, one given by $V'$, and one being simulated in $N$.
In this case the simulation graph (as defined above) would only have a single edge, from $V$ to $V'$. Yet there would
be two instances of us, evolving independently.
To complicate the situation further, we might also be running a simulation of ourselves.

In addition, any particular universe might be running
more than one simulation at once. Physically, this could occur by having one computer in that universe running multiple simulations
simultaneously, just like a laptop
runs multiple tasks simultaneously, by ``swapping''.  It could also occur by having 
multiple computers in the universe, all running simulations.\footnote{In terms of the formalism in this paper, the latter case would
mean that ``the'' computer $N$ of the universe is actually a set of multiple computers running independently, in parallel. The 
simulation lemma would apply directly, and the self-simulation lemma would also hold, where $\SSS(\Delta t)$
is the initial joint state of all of the computers in the universe. }

As an example, 
there might be a universe $V$ that is running simulations of multiple different universes, $V_1, V_2, \ldots$ simultaneously,
and as one runs down some of the simulation paths from some subset of those universes, $V_{i_1}, V_{i_2}, \ldots$, one eventually
converges at us. Similarly, there might be beings in a universe who are running us in their computer who themselves
are a simulation in a computer that \textit{we} are running. 
In such a case, the aliens would be a simulation running in a computer that they
themselves control, just ``one step removed''. The same would also be true of us of course. We would be
a means for the aliens to ``split'' their ontological status in two, while also being a means for us to split
\textit{our} ontological status in two.

This suggests an extension of the simulation graph, in which all the edges coming out of each node $V$ are labeled by 
one or more elements
of $\N$. The interpretation would be that for all $i \in \N$, the set of edges out of $V$ labeled $i$ are a 
maximal set of simulations that $V$ could be doing simultaneously. (``Simultaneous simulation'' could be formalized
by modifying \cref{def:sim} so that the same initial condition of the simulating computer, $n_0 \in N$, would
result in the computation of the future state of multiple evolution functions.)

\section{Implications of Rice's theorem for (self-)simulation}
\label{sec:elementary_properties}

Rice's theorem, discussed in \cref{app:recursion_theorem-rice_theorem}, has some interesting implications for both simulation
and self-simulation. To illustrate these, for simplicity, throughout this subsection I'm restricting attention to universes with
a countably infinite $W$ as well as a countably infinite $N$.

First, Rice's theorem tells us that the set of all computational universes $V'$ that can 
be simulated by a fixed universe $V$ is undecidable. 
More formally, fix some universe $V$ that simulates at least one other universe.
Define $A(V)$ as the collection of all TMs that compute the evolution function
of universes $V'$ that are simulated by $V$. Note that every TM that lies in $A(V)$ must be total, since all evolution
functions are. Therefore there are TMs that do not lie in $A(V)$ (e.g., all TMs that compute a partial function), 
as well as TMs that do lie in $V$ (by definition of $V$ and $A(V)$). Moreover, any 
two TMs that compute the exact same evolution function either both lie in $A(V)$ or both do not, i.e., membership in $A(V)$ does
not depend on \textit{how} the associated TM operates, only on the function it computes. 
Therefore by Rice's theorem, it is undecidable whether an arbitrary  
(total TM and associated) $V'$ is a member of $A(V)$. 


As a variant of this result, again fix $V$, and also fix some spaces $W', N'$. Define $B(V, (w'_0, n'_0), (w'_{\Delta t}, n'_{\Delta t}))$ 
as the collection of all TMs that compute the evolution function $g'$ of some universe $V' = W' \times N'$ with the following two properties.
First, $g'$ sends $(w'_0, n'_0)$ to $(w'_{\Delta t}, n'_{\Delta t})$. 
Second, $V$ simulates $V'$ \textit{for the specific initial condition} $(w'_0, n'_0)$
and the simulation time $\Delta t$. Again, it is undecidable whether an arbitrary TM lies in 
$B(V, (w'_0, n'_0), (w'_t, n'_t))$. 


Flipping things around, Rice's theorems tells us that
for any fixed universe $V'$, the set of all other universes $V$ that can simulate $V'$ is undecidable.
More formally, fix $V'$, and define the property of TMs that the function they compute is the evolution function
of a universe $V$ that simulates $V'$. Then it is undecidable whether an arbitrary such 
(TM and associated) $V$ has that property of simulating $V'$. An immediate consequence of this is that
the set of all pairs of universes $(V, V')$ such that $V$ simulates $V'$ is undecidable.



Rice's theorem also shows that:
\begin{enumerate}
\item The set of universes (and in particular evolution functions) that obey the RPCT is undecidable.
\item The set of universes that can simulate themselves is undecidable.
\item The set of universes $V$ that can simulate a universe $V' \ne V$ that can in turn simulate any universe at all is
undecidable.
\item The set of universes $V$ that can simulate a universe $V' \ne V$ that can in turn simulate $V$ is
undecidable.
\item Restrict attention to some set $\mathcal{V}$ of universes that can simulate themselves (e.g., because 
all the universes in $\mathcal{V}$ obey both
the PCT and the RPCT). The set of those universes in $\mathcal{V}$ that can simulate itself for 
all $\Delta t \in \N$ using some specific function $\SSS(\Delta t)$  is undecidable.
\item In particular, for any $k \in \{2, \ldots\}$, the set of universes in $\mathcal{V}$
that can simulate itself for an associated $\SSS(\Delta t)$ for all $\Delta t \le k$ but not for some $\Delta t > k$
is undecidable.
\item For any $k \in \{2, \ldots\}$, the set of universes in $\mathcal{V}$
that can simulate itself for an associated $\SSS(\Delta t)$ for all $\Delta t \ge k$ but not for some $\Delta t > k$
is undecidable.
\item The set of those universes in $\mathcal{V}$ that can simulate itself 
for a time-ordered $\SSS(\Delta t)$ is undecidable.
\end{enumerate}

There are some strange philosophical implications of these impossibility results, especially those that concern self-simulation. 
For example, it is possible that
we are in a universe $V$ that is simulating itself --- but only up to some future time, after which it is impossible
for the simulation to still be accurate. At that future time we would ``split'' into two versions of ourselves, which share
an identical past: the simulating version of us, and the simulated version of us. The impossibility results above say
that we can never be sure that this is not the case.

\section{Mathematical issues raised by the self-simulation and simulation lemmas}
\label{sec:math_implications}

There are many mathematical questions suggested by the self-simulation lemma
that I am not considering in this paper. Most obviously, I have not considered the computational complexity of finding 
$\SSS(\Delta t)$, and its dependence on $g$, $\Delta t$, $|W|$, etc. 

I also have not considered the relation between $\Delta t$ and
the physical time $\TT(\Delta t, w_0, n_0)$ at which a computer $N$ with initial state $n_0$ running a simulation finishes
its calculation of the state of $V$ at physical time $\Delta t$, when the initial state of the environment is
$w_0$. In particular, I have not considered how
the minimal value (over all $n_0$) of $\TT(\Delta t, w_0, n_0) / \Delta t$ might depend on $g$, $w_0$,
the value of $\Delta t$, etc. Associated questions, more in the spirit of computational complexity theory,
would involve the scaling of 
\eq{
\min_{n_0} \max_{w_0} \dfrac{\TT(\Delta t, w_0, n_0)}{\Delta t}
}
with $|W|$, for a fixed family of evolution functions $\{g_{|W|}\}$.

Next, define ``non-greedy nested self-simulation'' as the variant of nested self-simulation where for at least one $t_i$
in a simulation time sequence computed by the computer, $t_i$, the gap $t_{i+1} - t_i$ is not minimal.
Define the \textbf{density (of simulation times)} of an instance of nested self-simulation producing the simulation time sequence 
$\{t_1, t_2, \ldots\}$ for initial condition $v_0 = (w_0, n_0)$ as
\eq{
D(w_0, n_0) := \lim_{i \rightarrow \infty} \dfrac{i}{t_i}
} 
Note that this is well-defined even if the instance of nested self-simulation produced by $n_0$ isn't greedy.

One obvious question is what the properties of $g$ and $v_0$
need to be for the density $D(w_0, n_0)$ to be well defined. A related question is whether in scenarios where the density is 
well-defined, greedy self-simulation maximizes it.( \textit{A priori}, it could be that delaying some
simulation times $t_i$ allows denser subsequent simulation times.) A higher-level question is whether nested self-simulation,
greedy or otherwise, maximizes the density of simulation times over the set of all possible TMs.

There are also interesting computational complexity issues
concerning the three computable functions that define one universe's simulating another.
In particular, there are obvious extensions of the basic framework 
to concern not perfect simulation, but rather approximate simulation. This then immediately suggests
investigating variants of simulation graphs, defined by the 
the approximation complexity of (imperfect) simulation~\cite{arora2009computational}. For example, one
might consider simulation graphs where edges from $V$ to $V'$ must respect an upper bound on how fast the
function $\TT(\Delta t', w', n')$ must grow as a function of $\Delta t'$
in order for the resultant simulation of $V'$'s evolution to be at least a factor $\alpha$ within exact. As another
example, one could consider such approximation complexity in the computation of $\TT, \WW, \NN$ themselves,
rather than in the resultant behavior of $\TT$.

Similarly, we can consider the average-case complexity of all the issues arising in the
simulation framework. In this kind of approach we would again consider a variant of
simulation graphs, this time defined by requiring all edges from one universe (labeled as $V$) to a potentially different
universe (labeled as $V'$) in the simulation graph must obey
associated bounds. In this case though those bounds would concern the 
average-case behavior of $\TT(\Delta t', w', n')$ as a function of $\Delta t'$, or
of the resources needed to compute the functions $\TT, \WW, \NN$. Why might also want
to follow Levin in how precisely to define such ``average-case complexity''. 

There are many other open questions that involve slight variants of the framework introduced in this paper, in addition
to those discussed in the main text. For example, there are some ways to refine the definition of simulation
that might be worth pursuing. One of these is to define the ``time-minimal'' triple of simulation functions $(\TT, \WW, \NN)^{V,V'}$
used by $V$ to simulate $V'$ as the three such functions where
for all triples $(\Delta t', w'_0, n'_0)$, the associated time to complete the simulation,
$\TT^{V,V''}(\Delta t', w'_0, n'_0)$, is minimal. So intuitively, this triple of simulation functions
minimizes the time cost (in the complexity theory sense) for $V$ to perform the computation of the future state of $V'$.
Next define  
\eq{
\mathbb{T}^{V,V'}(\Delta t') := \max_{w'_0 \in W', n'_0 \in N'} \TT^{V,V'}(\Delta t', w'_0, n'_0)
}
This is the worst-possible time to complete the simulation using the fastest simulating computer.

It might be of interest to investigate the scaling of $\mathbb{T}^{V,V'}(\Delta t')$ as a function of $|V'|$, the size 
of the state space of $V'$. (In the case of self-simulation, $V = V'$, and we might instead investigate the
scaling of $\mathbb{T}^{V,V}(\Delta t')$ as a function of $|W|$.) In particular, to get a precise analogy with
the concept of time complexity in computational complexity theory, we might want to 
investigate how that scaling depends on both $g$ and $g'$.

Analogous issues would arise for a ``space minimal'' variant of simulation, involving the simulation function $\NN$ rather
than $\TT$. In particular, we could investigate how the scaling properties of the space-minimal
cost depends on $g, g'$. This would provide an analogy with
the concept of space complexity in computational complexity theory. In a similar way, it might be fruitful
to view time-minimal and / or space-minimal versions of the RPCT.


\section{Discussion}
\label{sec:discussion}

To my knowledge, \cref{lemma:sim} is the first fully formal
statement of what has been informally referred to in the literature as the ``simulation hypothesis''.
Going further, it is also the first formal derivation of a set of sufficient conditions
that ensure that the simulation hypothesis holds. \cref{lemma:2} then goes further, and establishes
sufficient conditions for a universe to have a computer that simulates itself, a possibility with
strange philosophical consequences. These lemmas also lead to many interesting questions that are
purely mathematical, e.g., concerning the simulation graph of universes simulating universes, the
minimal time delay in self-simulation, etc.

There have been informal discussions in the literature attacking the simulation hypothesis on 
the grounds that each successive level of simulation within simulation would necessarily be computationally
weaker than the one just above it. The idea is that due to this strict ``weakening of computational power'', there is a deepest possible level
of simulation, containing a species that is not computationally powerful enough to simulate any other species~\cite{carroll.simulation.hypothesis}.
The argument is made that this deepest level would only be a finite number of levels below the one we inhabit.

This argument has been criticized for not considering the possibility that the computational power of the successive levels
might \textit{asymptote} at some weakest amount of power. In this case there would actually be an infinite number of levels
below the one we inhabit, and none of them would be so weak as to be incapable of simulating yet a deeper level. The point
is moot however; the self-simulation lemma disproves the starting supposition of the argument, that  
``each successive level of simulation within simulation would necessarily be computationally
weaker than the one just above it''.

The self-simulation lemma also problematizes  --- perhaps fatally --- the whole idea of assigning a probability to the possibility that ``we are a simulation''. If in fact we are a \textit{self}-simulation, then we would be both the simulation, and the simulator.
Indeed, in some senses we would be an infinite number of simulations-within-simulations, all distinguishable
by how slowly they evolve, but in all others ways completely identical. That raises the obvious question
of how many instances of us we need to include in tallying up the number
of cases in which we're a simulation. Without answering that question, it is hard
to see how to calculate the probability of our being a simulation.

Numerous open issues concerning simulation and mathematics are discussed in the text. There
are also several open issues concerning simulation and the laws of physics as we currently understand them. In particular,
it might be worth investigating extensions of the analysis in this paper to concern quantum mechanical
and / or relativistic universes. One obvious question in this regard is whether the quantum no-cloning theorem means
that self-simulation could never arise (at the quantum level) in our universe. If so, that might point to ways for the
recursion theorem to be modified for quantum rather than classical computers.

\section*{Acknowledgments:} I would like to thank the Santa Fe Institute for support, and Aram Ebtekar for a close read.

\section*{Statements} I declare no conflicts of interest occurred in the preparation of this manuscript. I declare that
no ethical issues (e.g., concerning research on animals or human subjects) arose in the preparation of this manuscript.

\appendix

\section{Appendix A: Turing machines}
\label{app:TMs}


%
%

Perhaps the most famous class of computational machines are Turing 
machines~\cite{livi08,sipser2006introduction,arora2009computational,savage1998models}.
One reason for their fame is that it seems one can model any computational machine that
is constructable by humans as a Turing machine. A bit more formally, the \textit{Church-Turing
thesis} (CT) states that, ``A function on the natural numbers is computable by a human 
being mechanically following an algorithm, ignoring resource limitations, if and only if 
it is computable by a Turing machine.'' Note that it's not even clear whether this is a statement about the
physical world that could be true or false, or whether instead it is simply a definition, of what ``mechanically following an algorithm''
means~\cite{copeland2023church}. In any case,
the ``physical CT'' (PCT) modifies
the CT to hypothesize that the set of functions computable with Turing machines includes 
all functions that are computable using mechanical algorithmic procedures (i.e., those we humans can
implement) admissible by the laws of
physics~\cite{arrighi2012physical,piccinini2011physical,pour1982noncomputability,moore1990unpredictability}.

In earlier literature, the CT and PCT were both only always described semi-formally, or simply taken
as definitions of terms like ``calculable via a mechanical procedure'' or ``effectively computable'' (in contrast to the fully
formal definition given in \cref{sec:PCT}). Nonetheless, in part due to the CT thesis,
Turing machines form one of the keystones of the entire field
of computer science theory, and in particular of computational complexity~\cite{moore2011nature}.
For example, the famous Clay prize question of whether $\cs{P} = \cs{NP}$ --- widely considered one of the deepest and most profound open questions in mathematics ---
concerns the properties of Turing machines. As another example, the theory
of Turing machines is intimately related to deep results on the limitations of mathematics,
like G{\"o}del's incompleteness theorems, and seems to have broader
implications for other parts of philosophy
as well~\cite{aaronson2013philosophers}. Indeed, invoking the PCT, it has been argued that the foundations
of physics may be restricted by some of the properties of Turing machines~\cite{barrow2011godel,aaro05}.

Along these lines, some authors have suggested that the foundations of statistical
physics should be modified to account for the properties of Turing machines, e.g.,
by adding terms to the definition of entropy. After all, given the
CT, one might argue that the probability distributions at the heart of
statistical physics are distributions ``stored in the
mind'' of the human being analyzing a given statistical physical system (i.e., 
of a human being running a particular algorithm to compute a property of a given
system). Accordingly, so goes the argument, the costs
of generating, storing, and transforming the minimal specifications of
the distributions concerning a statistical physics system should be included 
in one's thermodynamic analysis of those changes in the distribution of states of the system.  See~\cite{caves1990entropy,caves1993information,zure89b}. 

There are many different definitions of Turing machines that are computationally equivalent
to one another, in that
any computation that can be done with one type of Turing machine can be done with the other.
It also means that the ``scaling function'' of one type of Turing machine, 
mapping the size of a computation to the minimal amount of resources needed to 
perform that computation by that type of Turing machine, is at most a polynomial
function of the scaling function of any other type of Turing machine. (See for example the
relation between the scaling functions of single-tape and multi-tape
Turing machines~\cite{arora2009computational}.) The following definition will be useful
for our purposes, even though it is more complicated than strictly needed:

\begin{definition}
A \textbf{Turing machine} (TM) is a 7-tuple $(R,\Lambda ,b,v,r^\varnothing,r^A,\rho)$ where:

\begin{enumerate}
\item $R$ is a finite set of \textbf{computational states};
\item $\Lambda$ is a finite \textbf{alphabet} containing at least three symbols;
\item $b \in \Lambda$ is a special \textbf{blank} symbol;
\item $v \in \Z$ is a \textbf{pointer};
\item $r^\varnothing \in R$ is the \textbf{start state};
\item $r^A \in R$ is the \textbf{accept state}; and
\item $\rho : R \times \Z \times \Lambda^\infty \rightarrow 
R \times \Z \times \Lambda^\infty$ is the \textbf{update function}.
It is required that for all triples $(r, v, T)$, that if we write
$(r', v', T') = \rho(r, v, T)$, then $v'$ does not differ by more than $1$
from $v$, and the vector $T'$ is identical to the vectors $T$ for all components
with the possible exception of the component with index $v$;\footnote{Technically 
the update function only needs to be defined on the ``finitary'' subset of $\R \times \Z 
\times \Lambda^\infty$, namely, those elements of $\R \times \Z 
\times \Lambda^\infty$ for which the tape contents has a non-blank value in only finitely many positions.}
\end{enumerate}
\label{def:tm}
\end{definition}

$r^A$ is often called the ``halt state'' of the TM rather than the accept state. 
(In some alternative, computationally equivalent definitions of TMs, 
there is a set of multiple accept states rather than a single accept state, but for simplicity I do not
consider them here.) $\rho$ is sometimes called the ``transition function'' of the TM.
We sometimes refer to $R$ as the states of the ``head'' of the TM,
and refer to the third argument of $\rho$ as a \textbf{tape}, writing a
value of the tape (i.e., semi-infinite string of elements of the alphabet) as $\lambda$.
The set of triples that are possible arguments to the update function 
of a given TM are sometimes called the set of \textbf{instantaneous descriptions}
(IDs) of the TM. (These are sometimes instead referred to as ``configurations''.)
Note that as an alternative to Def.~\ref{def:tm}, we
could define the update function of any TM as a map over an associated space of IDs.

Any TM $(R,\Lambda ,b,v,r^\varnothing, r^A, \rho)$ starts with $r = r^\varnothing$, the counter
set to a specific initial value (e.g, $0$), and with $\lambda$
consisting of a finite contiguous set of non-blank symbols, with
all other symbols equal to $b$. The TM operates by iteratively
applying $\rho$, if and until the computational state falls in $r^A$, at
which time the process stops, i.e., any ID with the head in the halt state is a
fixed point of $\rho$.

If running a TM on a given initial state of the tape results in the TM eventually halting,
the largest blank-delimited string that contains the position of the pointer 
when the TM halts is called the TM's \textbf{output}. The initial
state of $\lambda$ (excluding the blanks) is sometimes called the associated 
\textbf{input}, or \textbf{program}. (However,
the reader should be warned that the term ``program'' has been used by some physicists to
mean specifically the shortest input to a TM that results in it computing
a given output.) We also say that the TM \textbf{computes} an output
from an input.  In general though, there will be inputs for which the TM never halts. 
The set of all those inputs to a TM that cause it to eventually
halt is called its \textbf{halting set}. We write the output of a TM $T$ run on an input
$x$ that lies in its halting set as $T(x)$. 

Write the set of non-blank symbols of $\Lambda$ as $\hat{\Lambda}$.
Every $\lambda$ on the tape of a TM that it might have during a computation in which it halts is a finite string of elements in $\Lambda$
delimited by an infinite
string of blanks. Accordingly, wolog we often refer to  the state space of the tape is $\Lambda^*$, with
the trailing infinite string of blanks implicit. Note that $\Lambda^*$ is countably infinite, in contrast to $\Lambda^\infty$.

If a function is undefined for some elements in its domain, it is called a \textbf{partial} function.
Otherwise it is a \textbf{total} function. In particular 
if a TM $T$ does not halt for some of its inputs, so its halting set is a proper subset of its domain,
then the map from its domain to outputs is a {partial function}, and if instead its halting set is
its entire domain, it is a total function.

We say that a total function $f$ from $(\Lambda \setminus \{b\})^*$ to itself
is \textbf{recursive}, or \textbf{(total) computable}, if there is a TM with input alphabet $\Lambda$ such that for all 
$x \in (\Lambda \setminus \{b\})^*$, the TM computes $f(x)$. If $f$ is instead a partial function,
then we say it is \textbf{partial recursive} (partial computable, resp.) if there is a TM with input alphabet $\Lambda$
that computes $f(x)$ for all $x$ for which $f(x)$ is defined, and does not halt for any other $x$. (The 
reader should be warned that in the literature, the term ``computable'' is sometimes taken to mean
partial computable rather than total computable --- and in some articles it is sometimes taken to mean either
total computable or partial computable, depending on the context.) 

An important special case is when 
the image of $f$ is just $\B$, so that for all $s \in (\Lambda \setminus \{b\})^*$, $f(s)$ is just a single bit.
In this special case, we say that the set of all $s : f(s) = 1$ is \textbf{decidable} if $f$ is computable.

Famously, Turing showed that there are total functions that are not recursive. In light of the CT,
this result is arguably one of the deepest philosophical truths concerning fundamental 
limitations on human capabilities ever discovered. (See~\cite{copeland2023church,raattkainen2005philosophical,piccinini2010computation,piccinini2011physical}.)

As mentioned, there are many variants of the definition of TMs provided above. In one
particularly popular variant the single tape in \cref{def:tm}
is replaced by multiple tapes. Typically one of
those tapes contains the input, one contains the TM's output (if and) when the TM
halts, and there are one or more intermediate ``work tapes'' that are
in essence used as scratch pads. The advantage of using this more complicated
variant of TMs is that it is often easier to prove theorems for such machines
than for single-tape TMs. However, there is no difference in
their computational power. More precisely, one can transform any single-tape TM
into an equivalent multi-tape TM (i.e., one that computes the same partial function),
as well as vice-versa~\cite{arora2009computational,livi08,sipser2006introduction}.

To motivate an important example of such multi-tape TMs, 
suppose we have two strings $s^1$ and $s^2$ both contained in the set $(\Lambda \setminus \{b\})^*$ where $s^1$ is a proper prefix of $s^2$. 
If we run the TM on $s^1$, it can detect when it gets to the end of its input, by
noting that the following symbol on the tape is a blank. Therefore, it can
behave differently after having reached the end of $s^1$ from how it behaves
when it reaches the end of the first $\ell(s^1)$ bits in $s^2$. As a result,
it may be that both of those input strings are in its halting set, but result
in different outputs.

A \textbf{prefix (free) TM} is one in which this can never happen:
there is no string in its halting set that is a proper prefix of another string in its halting 
set. The easiest way to construct such TMs is to have a multi-tape TM with a single
read-only input tape whose head cannot reverse, and a write-only output
tape whose head cannot reverse, together with an arbitrary number of work tapes
with no such restrictions.\footnote{It is not trivial to construct prefix single-tape TMs directly.
For that reason it is common to use prefix three-tape TMs, in
which there is a separate input tape that can only be read from, output tape that
can only be written to, and work tape that can be both read from and written
to. To ensure that the TM is prefix, we require that the head cannot ever back up on
the input tape to reread earlier input bits, nor can it ever back up on the output
tape, to overwrite earlier output bits. To construct a single-tape prefix TM,
we can start with some such three-tape prefix TM and transform
it into an equivalent single-tape prefix TM, using any of the conventional techniques for transforming between
single-tape and multi-tape TMs.} I will often implicitly assume that any TM being discussed
is such a multi-tape prefix TM.

Returning to the TM variant defined in \cref{def:tm}, one of the most important results in CS theory
is that the number of TMs is countably infinite. This means that we can index the set of
all TMs with $\N$, i.e., we can write the set of TMs as $\{T^k : k \in \N\}$. It also 
means that there exist \textbf{universal Turing machines} (UTMs), $U$, which can be used
to emulate an arbitrary other TM $T^k$ for any $k$. More precisely, 
we define a UTM $U$ as one with the property that
for any other TM $T$, there is an invertible map $f$ from the set of possible 
states of the input tape of $T$ into the set of possible states of the input tape of $U$ with the following
properties: Both $f$
and $f^{-1}$ are computable, and if we apply $f$ to any input string $\sigma'$ of $T$ to construct an input string $\sigma$
of $U$, then:
\begin{enumerate}
\item $U$ run on its input $\sigma$ halts iff $T$ run on its input $\sigma'$ halts;
\item If $U$ run on $\sigma$ halts, and we apply $f^{-1}$ to the resultant output of $U$,
	we get the output computed by $T$ if it is run on $\sigma'$.
\end{enumerate}  

As is standard, I fix some set of (prefix free) encodings of all tuples of finite bit strings, $\langle . \rangle$, $\langle ., .\rangle$... 
In particular, in general the input to a UTM is encoded as $\langle k, x\rangle$ if it is emulating TM $T^k$ running on input string $x$.
However, sometimes for clarity of presentation I will leave the angle brackets implicit, and simply write the
UTM $U$ operating on input $\langle k, x \rangle$ as $U(k, x)$. In addition, as shorthand, if $x$ is a vector whose components
are all bit strings, I will write the encoded version of all of its components as $\langle x \rangle$.

Intuitively, the proof of the existence of UTMs just means that there exists programming languages which are ``(computationally) universal'',
in that we can use them to implement any desired program in any other language, after
appropriate translation of that program from that other
language. This universality leads to a very important concept:

\begin{definition}
The \textbf{Kolmogorov complexity} of a UTM $U$ to compute a string $\sigma \in \Lambda^*$
is the length of the shortest input string $s$ such that $U$ computes $\sigma$ from $s$.
\end{definition}
\noindent 
Intuitively, (output) strings that have low Kolmogorov complexity for some specific UTM $U$ are those with 
short, simple programs in the language of $U$. For example, in all common (universal) programming
languages (e.g., \textit{C, Python, Java}, etc.), 
the first $ m$ digits of $\pi$ have low Kolmogorov complexity, since those
digits can be generated using a relatively short program.  
Strings that have high (Kolmogorov) complexity are sometimes referred to as
``incompressible''. These strings have no patterns in them that can be generated by
a simple program. As a result,
it is often argued that the expression ``random string'' should only be used for strings that 
are incompressible.

We can use the Kolmogorov complexity of prefix TMs to define many associated quantities, which
are related to one another the same way that various kinds of Shannon entropy
are related to one another. For example, loosely speaking, the
conditional Kolmogorov complexity of string $s$ conditioned on string $s'$,
written as $K(s \mid s')$, is the length of the shortest string x such
that if the TM starts with an input string given by the concatenation
$xs'$, then it computes $s$ and halts. If we restrict attention
to prefix-free TMs, then for all strings $x, y \in \Lambda^*$, we have~\cite{livi08}
\eq{
K(x, y) \le K(x) + K(x \mid y) + O(1) \le K(x) + K(y) + O(1)
\label{eq:kolmogorov_conditional}
}
(where ``$O(1)$'' means a term that is independent of both $x$ and $y$).
Indeed, in a certain technical sense, the expected value of $K(x)$ under any
distribution $P(x \in \Lambda^*)$ equals the Shannon entropy of $P$. (See~\cite{livi08}.)

Formally speaking, the set $\B^*$ is a Cantor set. A convenient
probability measure on this Cantor set, sometimes called the \textbf{fair-coin measure}, 
is defined so that for any binary string $x$ the set of sequences that begin with $\sigma$ has measure $2^{-|\sigma|}$.
Loosely speaking, the fair-coin measure of a prefix TM $T$ is the probability distribution 
over the strings in $T$'s halting set generated by IID ``tossing a coin'' 
to generate those strings, in a Bernoulli process, and then normalizing.\footnote{Kraft's 
inequality guarantees that since the set of strings in the halting set is a prefix-free
set, the sum over all its elements of their probabilities cannot exceed $1$, and
so it can be normalized. However, in general that normalization constant
is uncomputable, as discussed below. Also, in many contexts we can actually
assign arbitrary non-zero probabilities to the strings outside the halting set,
so long as the overall distribution is still normalizable. See~\cite{livi08}.} 
So any string $\sigma$ in the halting set
has probability $2^{-|\sigma|} / \Omega$ under the fair-coin prior, where
$\Omega$ is the normalization constant for the TM in question.

The fair-coin prior provides a simple Bayesian interpretation of Kolmogorov
complexity: Under that
prior, the Kolmogorov complexity of any string $\sigma$ for any prefix TM $T$ is just
(the log of) the maximum a posterior (MAP) probability that any string $\sigma'$ in the halting
set of $T$ was the \textit{input} to $T$, conditioned on $\sigma$ being
the \textit{output} of that TM. (Strictly speaking, this result is only
true up to an additive constant, given by the log of the normalization
constant of the fair-coin prior for $T$.)
%

The normalization constant $\Omega$ for any fixed prefix UTM, sometimes called ``Chaitin's Omega'',
has some extraordinary properties. For example, the successive digits of $\Omega$
provide the answers to \textit{all} well-posed mathematical problems. So if we
knew Chaitin's Omega for some particular prefix UTM, we could answer every problem in mathematics.
Alas, the value of $\Omega$ for any prefix UTM $U$ 
cannot be computed by any TM (either $U$ or some other one).
So under the CT, we cannot calculate $\Omega$. 
(See also~\cite{baez2012algorithmic} for a discussion of a
``statistical physics'' interpretation of $\Omega$ that results if we view the fair-coin prior
as a Boltzmann distribution for an appropriate Hamiltonian, 
so that $\Omega$ plays the role of a partition function.)

It is now conventional to analyze Kolmogorov complexity using prefix UTMs, with the fair-coin
prior, since this removes some undesirable technical properties that Kolmogorov complexity has
for more general TMs and priors. Reflecting this, all analyses in the physics
community that concern TMs assume prefix UTMs. 
(See~\cite{livi08} for a discussion of the extraordinary properties of such UTMs.)

Interestingly, for all their computational power, there are some surprising ways
in which TMs are \textit{weaker} than the other computational machines introduced above.
For example, there are an infinite number of TMs that are more powerful than any given circuit, i.e., 
given any circuit $C$, there are an infinite number of TMs that compute the same function as $C$.
Indeed, any single UTM is more powerful than \textit{every} circuit in this sense. On the other hand,
it turns out that there are circuit \textit{families} that are more powerful than any single TM.
In particular, there are circuit families that can solve the halting problem~\cite{arora2009computational}.
 
I end this appendix with some terminological comments and definitions that will
be useful in the main text. It is conventional when dealing with Turing machines to implicitly
assume some invertible map $h(.)$ from $\Z$ into $\Lambda^*$. Given
such a map $h(.)$, we can exploit it to implicitly assume an additional invertible
map taking $\mathbb{Q}$ into $\Lambda^*$, e.g., by uniquely expressing any rational
number as one product of primes, $a$, divided by a product of different primes, $b$;
invertibly mapping those two products of primes into the single integer $2^a 3^b$; and then evaluating
$Rh(2^a 3^b)$.
Using these definitions, we say that a real number $z$ is \textbf{computable}
iff there is a recursive function $f$ mapping rational numbers to rational
numbers such that for all rational-valued accuracies $\epsilon > 0$,
$|f(\epsilon) - z| < \epsilon$. We define computable functions from $\mathbb{Q} \rightarrow \R$
similarly.

%
%
%

\section{Appendix B: The recursion theorem and Rice's theorem}
\label{app:recursion_theorem-rice_theorem}

Any reader not already familiar with the theory of Turing machines should read \cref{app:TMs} before
this appendix.

\subsection{The recursion theorem}

Kleene's second recursion theorem~\cite{sipser1996introduction,kleene1952introduction,moschovakis2010kleene,avigad2007computability}
can be stated as follows:

\begin{theorem}
For any partial computable function $Q(x, y)$ there is a Turing machine with index $e$ 
such that $T^e(x) = Q(x, e)$ for all $x$.
\end{theorem}

An elegant proof of an extended version of Kleene's second recursion theorem can be found in~\cite{moschovakis2010kleene}.
In terms of the notation in this paper, it proceeds as follow:

\begin{proof}
For any TM of the form $M(y, x)$ taking two arguments, define $\llbracket{M(y, x)}\rrbracket_x$ as the index $e$
such that $T^e(x) = M(y, x)$ for all $x$, with $y$ fixed. 

Our goal is to find a way to satisfy this equation for $e = y$. 
In other words, we wish to find a $y$ so that
\eq{
T^{\llbracket{M(y, x)}\rrbracket_x}(x) = M(y, x)
}

Using this notation, define
\eq{
S(t) &:= \llbracket{T^t(t, x)}\rrbracket_x
\label{eq:rec_thm_proof_1}
}
Note that $S(.)$ is a total computable function. 
Using \cref{eq:rec_thm_proof_1}, choose one out of the infinite set of indices $k$ such that
\eq{
T^k(t, x) = Q(x, S(t))
\label{eq:rec_thm_proof_2}
}
(Such an index must exist since the RHS of \cref{eq:rec_thm_proof_2} is a partial computable function, and
by running over $k$ the LHS runs over all TMs.)

Finally, set $t = k$ in \cref{eq:rec_thm_proof_2} and then plug in \cref{eq:rec_thm_proof_1}. The proof is completed
by choosing
\eq{
e &:= \llbracket{T^k(k, x)}\rrbracket_x
}
\end{proof}

In computer science theory, Kleene's second recursion theorem is just called ``the recursion theorem''.
It has played an extremely important role in computer science theory. For example
it provides the underlying formal justification for Von Neumann's universal constructor, which in turn
was extremely important for understanding the foundations of biology. More prosaically, it
provides the formal justification for why computer viruses are possible (assuming we use computers
that are Turing complete).

An important special case of the theorem is where $Q(., .)$ is a \textit{total} computable function.
In this case the theorem says that that there must be an $e$ such that $T^e(x) = Q(x, e)$ for all $x$,
and therefore $T^e(x)$ is also total computable. So as a special case of the recursion
theorem we can restrict to total computable functions $Q(., )$ and
total TM's $T^e$. I call this special case the ``total recursion theorem'' in the main text.

%

As an aside, the recursion theorem has elicited some pointed commentary. For example,~\cite{moschovakis2010kleene} writes that
\begin{quote}
``The proof has always seemed too short and tricky, and some considerable effort has gone into explaining how one discovers it short of
``fiddling around'' ... Some of his students asked Kleene about it once, and his (perhaps facetious) response
was that he just ``fiddled around''' --- but his fiddling may have been informed by similar results in the untyped $\lambda$-calculus.''
\end{quote}

\noindent Another interesting comment was made by Juris Hartmanis~\cite{gaulle1985upson}:

\begin{quote}
``The recursion theorem is just like tennis. Unless you're exposed to it at age five, you'll never become world class.''
\end{quote}

\noindent Interestingly, Hartmanis didn’t encounter the recursion theorem until he was in his $20$’s --- and yet 
despite the implications of his comment, he went on to win the Turing award.

%
%

\subsection{Rice's theorem}

Rice's theorem is an extremely powerful theorem about computability which can be proven from Kleene's
second recursion theorem. (This is shown in the wikipedia entry on Rice's theorem, for example.)
Perhaps the simplest way to state it is the following:

\begin{quote}
Let $G$ be any non-empty set of partial computable functions (e.g., represented as a set
of bit strings that encode the TMs that compute those functions). 
Suppose I can design an algorithm that correctly determines whether \underline{any} specific partial computable function $f$ 
lies in $G$,
i.e., suppose that membership in $G$ is decidable. 
Rice's theorem says that if this is the case, then $G$
must be the set of \textit{all} partial computable functions, 
i.e., our algorithm must always produce the output, ``yes''. 
\end{quote}

\noindent Intuitively, fix some property $G$ of the functions that can be computed by TMs, and suppose
we design an algorithm to decide whether the function computed by an arbitrary TM lies in $G$.  
Rice's theorem tells us that either $G$ is
the set of \textit{all} such functions, or there are some TMs that our algorithm
fails on. 

So if there is any TM $T$ that (partially) computes a function
that is not in $G$, then there must be a TM $T'$ such that our algorithm cannot tell us whether
the function that $T'$ computes lies in $G$.

An important special case is where $G$ is restricted to a set of partial functions each of which outputs a single bit. In this case 
Rice's theorem concerns the decidability of sets of languages.

\section{Appendix C: Why $\Delta t$ is not a physical variable}
\label{sec:why_time_not_in_V}

Recall that as mentioned in~\cref{sec:features_self_sim}, $\Delta t$ is not a physical variable, but rather
it is a parameter of the evolution function. At first one might think that it makes
more sense to have $\Delta t$ be a physical variable in the universe, fixed in the value $v_0$. 
The idea would be to design the framework so that if we
change the value of this physical $\Delta t$ from some $t_1$ to some $t_2 \ne t_1$, without changing any of the rest of the universe,
then $V$ would simulate $V'$ for $t_2$ iterations into the future rather than $t_1$.
iterations In addition, in this alternative approach $g$ would
only be an explicit function of $w_0$ and $n_0$, and not of $\Delta t$.

This would be particularly problematic in the case of self-simulation though. In general, we are interested
in evolving an arbitrary initial state of the universe an arbitrary time into the future. That initial state and that
time into the future that interests us are completely independent. In particular, 
we might be interested in evolving the initial state of the universe to a future time that differs
from whatever value the variable $\Delta t$ specified in $v_0$ might have. So it would seem that
we need \textit{two} times into the future to be specified in this alternative framework.

There are also more formal problems with this alternative approach. 
Suppose that $\Delta t$ were specified as the initial value of a component of $w$, the physical variable giving the universe external 
to the computer. In this approach, there would be no way to have $\WW\left(\Delta t', w'_0, n'_0\right) = w'_0$
(in order to get free simulation), while also having $\TT\left(\Delta t', w'_0, n'_0\right) \ne \Delta t'$.
Yet as described below, in fact it is impossible to have $\TT\left(\Delta t', w'_0, n'_0\right) = \Delta t'$
for all $\Delta t'$ --- the pristine RPCT could not hold if we imposed that requirement.

On the other hand, suppose that rather than having $\Delta t$ be specified as a component of $w_0$, we had the
initial state of the computer be some string $\anglebracket{p_0, \Delta t}$, where we want to view $p_0$ as a fixed
``simulation program'' that would run on the computer, taking $\Delta t$ as input.
In other words, suppose that
$\Delta t$ were always specified as part of the initial state of the computer, $n$, and $g$ did not involve
$\Delta t$ directly. In this case, because $g$ itself does not vary with $\Delta t$ the recursion theorem
would not just fix the initial simulation program we want the computer to run, but also the time $\Delta t$
into the future we are running it. 
We would not be able to simulate the evolution of the universe to an 
arbitrary time in the future.\footnote{A subtlety is that the recursion theorem can in general be satisfied
by more than one $n^*$ --- by an infinite number in fact. It is not clear though that there is a way to exploit
this flexibility so that there is at least one $n^*$ that satisfies the recursion theorem for all $\Delta t$. So
for simplicity, this possibility is not considered in this paper.}

Ultimately, the way we are getting around these problems in the framework I've adopted is by having  $\Delta t$ just be a parameter
of the evolution function, specifying how far into the future we want to simulate the evolution of the physical
universe. and not a physical variable. So for example
it does not need to be reproduced
as the output of the computer if the cosmological universe is to simulate its own future.)
A consequence of this workaround is that we need to hard-code $n^*$ into the
initial state of the program in a way that depends on $\Delta t$.\footnote{Indeed, the way the
derivation of the self-simulation lemma uses the recursion theorem can be viewed as a special case of
the generalized parameter-dependent form of the recursion theorem given in~\cite{moschovakis2010kleene}. 
The derivation of the self-simulation lemma uses that generalized form for the special case where the space
of parameters is single-dimensional. So in the notation of that paper, here $m = 1$.}

\section{Appendix D: Subtleties with the model in \cref{sec:example_V_our_universe}}
\label{app:CAs_not_TMs}

\cref{sec:example_V_our_universe}
presented an example of how a portion of our actual universe could implement the purely formal model of universe $V$ 
given in \cref{sec:framework}. In that example the computer was implemented as 
a UTM, replete with tapes, etc.. This might seem a particularly awkward way of 
modeling a portion of our cosmological universe whose dynamics is computationally universal. 
After all, our universe ``runs in parallel'', whereas a UTM is a serial system.
This means that to implement such a UTM would require copious
use of energy barriers and the like, to prevent the parallel nature of Hamilton's equations from ``leaking through''.

Given that, it might seem more reasonable to use an infinite one-dimensional cellular 
automata (CA~\cite{wolfram1984cellular,israeli2004computational,mitc96,codd68}) running a computationally universal
rule, as a model of a universal computer that is purely parallel. Arguing against this though, one might object that such a CA
performs an infinite number of operations in parallel in each iteration. A single conventional TM, operating on one cell
per tape in each iteration, could not execute any such single iteration of a CA in finite time. So such an infinite CA
is doing something that is beyond the ability of a Turing machine.

Despite this though, in point of fact one-dimensional CA
are \textit{not} viewed in the literature as more powerful than TMs. The reason is that
that infinite number of parallel operations done by a CA does not provide it the ability
to compute functions that cannot be computed by Turing machines. (E.g., one cannot use a one-dimensional
CA to solve the halting problem.) Formally, this discrepancy is resolved by working with arbitrarily large --- but finite --- sub-strings
of an infinite one-dimensional CA.


Another subtlety is that in the example in \cref{sec:example_V_our_universe}, the first thing that happens when
$V$ evolves is that $w_0$ is copied onto the input tape. That single operation could take an 
arbitrarily large number of iterations of the UTM, depending on the size of $W$. This would
require  adding some large constant dependent on $|W|$ to many of the calculations in this paper. 
To avoid having to consider this technical issue, it might be convenient to implicitly modify 
$V$ so that $N$ evolves as a conventional UTM that is augmented with a special instruction. That
special instruction copies an arbitrarily number of bits from $W$ into $N$ in a single iteration. Similarly,
it might be convenient to include a special instruction that copies an 
arbitrarily large number of bits from one portion of $N$ to another portion. Whether to consider such a modified
$V$ or not is really just a matter of taste.

\bibliographystyle{amsplain}
\bibliography{/Users/davidwolpert/Dropbox/BIB/refs,/Users/davidwolpert/Dropbox/BIB/refs.main.1.BIB.DIR,/Users/davidwolpert/Dropbox/LANDAUER.Shared.2016/thermo_refs.main,/Users/davidwolpert/Dropbox/LANDAUER.Shared.2016/thermo_refs_2}

\newcommand{\arXiv}[2]{\href{http://arxiv.org/abs/#1}{arXiv:#1 #2}}
\providecommand{\bysame}{\leavevmode\hbox to3em{\hrulefill}\thinspace}
\providecommand{\MR}{\relax\ifhmode\unskip\space\fi MR }
\providecommand{\MRhref}[2]{%
  \href{http://www.ams.org/mathscinet-getitem?mr=#1}{#2}
}
\providecommand{\href}[2]{#2}
\begin{thebibliography}{10}

\bibitem{aaro05}
S.~Aaronson, \emph{{NP}-complete problems and physical reality},
  quant-ph/0502072, 2005.

\bibitem{aaronson2013quantum}
Scott Aaronson, \emph{Quantum computing since democritus}, Cambridge University
  Press, 2013.

\bibitem{aaronson2013philosophers}
\bysame, \emph{Why philosophers should care about computational complexity},
  Computability: Turing, G{\"o}del, Church, and Beyond (2013), 261--327.

\bibitem{ainsworth2010ontic}
Peter~Mark Ainsworth, \emph{What is ontic structural realism?}, Studies in
  History and Philosophy of Science Part B: Studies in History and Philosophy
  of Modern Physics \textbf{41} (2010), no.~1, 50--57.

\bibitem{arora2009computational}
Sanjeev Arora and Boaz Barak, \emph{Computational complexity: a modern
  approach}, Cambridge University Press, 2009.

\bibitem{arrighi2012physical}
Pablo Arrighi and Gilles Dowek, \emph{The physical church-turing thesis and the
  principles of quantum theory}, International Journal of Foundations of
  Computer Science \textbf{23} (2012), no.~05, 1131--1145.

\bibitem{avigad2007computability}
Jeremy Avigad, \emph{Computability and incompleteness},  (2007).

\bibitem{baez2012algorithmic}
John Baez and Mike Stay, \emph{Algorithmic thermodynamics}, Mathematical
  Structures in Computer Science \textbf{22} (2012), no.~05, 771--787.

\bibitem{barrow2007living}
John~D Barrow, \emph{Living in a simulated universe}, na, 2007.

\bibitem{barrow2011godel}
\bysame, \emph{Godel and physics}, Kurt G{\"o}del and the Foundations of
  Mathematics: Horizons of Truth (2011), 255.

\bibitem{beane2014constraints}
Silas~R Beane, Zohreh Davoudi, and Martin J.~Savage, \emph{Constraints on the
  universe as a numerical simulation}, The European Physical Journal A
  \textbf{50} (2014), no.~9, 148.

\bibitem{bostrom2003we}
Nick Bostrom, \emph{Are we living in a computer simulation?}, The Philosophical
  Quarterly \textbf{53} (2003), no.~211, 243--255.

\bibitem{campbell2017testing}
Tom Campbell, Houman Owhadi, Joe Sauvageau, and David Watkinson, \emph{On
  testing the simulation theory}, arXiv preprint arXiv:1703.00058 (2017).

\bibitem{miranda-cardona2021constructing}
Robert Cardona, Eva Miranda, Daniel Peralta-Salas, and Francisco Presas,
  \emph{Constructing turing complete euler flows in dimension 3}, Proceedings
  of the National Academy of Sciences \textbf{118} (2021), no.~19, e2026818118.

\bibitem{carroll.simulation.hypothesis}
Sean Carroll, \emph{Maybe we do not live in a simulation: The resolution
  conundrum}, Blog post, 2016.

\bibitem{carroll_wilczek_simulation_universe_2021}
Sean Carroll and Frank Wilczek, \emph{Frank wilczek on the present and future
  of fundamental physics},
  \url{https://www.preposterousuniverse.com/podcast/2021/01/18/130-frank-wilczek-on-the-present-and-future-of-fundamental-physics}.

\bibitem{caves1990entropy}
Carlton~M Caves, \emph{Entropy and information: How much information is needed
  to assign a probability}, Complexity, Entropy and the Physics of Information
  (1990), 91--115.

\bibitem{caves1993information}
\bysame, \emph{Information and entropy}, Physical Review E \textbf{47} (1993),
  no.~6, 4010.

\bibitem{chaitin2011goedel}
Gregory Chaitin, Francisco~A Doria, and Newton~CA Da~Costa, \emph{Goedel's way:
  Exploits into an undecidable world}, CRC Press, 2011.

\bibitem{chalmers2022reality+}
David~J Chalmers, \emph{Reality+: Virtual worlds and the problems of
  philosophy},  (2022).

\bibitem{conitzer2019puzzle}
Vincent Conitzer, \emph{A puzzle about further facts}, Erkenntnis \textbf{84}
  (2019), no.~3, 727--739.

\bibitem{cooper2017computability}
S~Barry Cooper, \emph{Computability theory}, Chapman and Hall/CRC, 2017.

\bibitem{copeland2023church}
B~Jack Copeland, \emph{The church-turing thesis}, The Stanford Encyclopedia of
  Philosophy (2023).

\bibitem{copeland2018church}
B~Jack Copeland and Oron Shagrir, \emph{The church-turing thesis: logical limit
  or breachable barrier?}, Communications of the ACM \textbf{62} (2018), no.~1,
  66--74.

\bibitem{cubitt2015undecidability}
Toby~S Cubitt, David Perez-Garcia, and Michael~M Wolf, \emph{Undecidability of
  the spectral gap}, Nature \textbf{528} (2015), no.~7581, 207--211.

\bibitem{dainton2012singularities}
Barry Dainton, \emph{On singularities and simulations}, Journal of
  Consciousness Studies \textbf{19} (2012), no.~1-2, 42--85.

\bibitem{codd68}
Codd~E. F., \emph{Cellular automata}, Academic Press, New York, 1968.

\bibitem{french2010defence}
Steven French and James Ladyman, \emph{In defence of ontic structural realism},
  Scientific structuralism, Springer, 2010, pp.~25--42.

\bibitem{gaulle1985upson}
Al~Gaulle, \emph{Upson's familiar quotations. (1984-1985)}, Tech. report,
  Cornell University, 1985.

\bibitem{hajek1979arithmetical}
Petr H{\'a}jek, \emph{Arithmetical hierarchy and complexity of computation},
  Theoretical Computer Science \textbf{8} (1979), no.~2, 227--237.

\bibitem{hamieh2021simulation}
Salah Hamieh, \emph{On the simulation hypothesis and its implications}, Journal
  of Modern Physics \textbf{12} (2021), no.~5, 541--551.

\bibitem{hanson2001live}
Robin Hanson, \emph{How to live in a simulation}, Journal of Evolution and
  Technology \textbf{7} (2001), no.~1, 3--13.

\bibitem{israeli2004computational}
N.~Israeli and N.~Goldenfeld, \emph{Computational irreducibility and the
  predictability of complex physical systems}, Physical review letters
  \textbf{92} (2004).

\bibitem{kipping2020objective}
David Kipping, \emph{An objective bayesian analysis of life's early start and
  our late arrival}, Proceedings of the National Academy of Sciences
  \textbf{117} (2020), no.~22, 11995--12003.

\bibitem{kleene1952introduction}
Stephen~Cole Kleene, NG~De~Bruijn, J~de~Groot, and Adriaan~Cornelis Zaanen,
  \emph{Introduction to metamathematics}, vol. 483, van Nostrand New York,
  1952.

\bibitem{livi08}
M.~Li and Vitanyi P., \emph{An introduction to kolmogorov complexity and its
  applications}, Springer, 2008.

\bibitem{lim2022we}
Abraham Lim, \emph{Why we are not living in a computer simulation},
  International journal for the study of skepticism \textbf{12} (2022), no.~4,
  331--351.

\bibitem{lloyd2013universe}
Seth Lloyd, \emph{The universe as quantum computer}, A Computable Universe:
  Understanding and exploring Nature as computation (2013), 567--581.

\bibitem{mccabe2006structural}
Gordon McCabe, \emph{Structural realism and the mind},  (2006).

\bibitem{mccabe2005possible}
Gordon McCabe et~al., \emph{Possible physical universes}, Zagadnienia
  Filozoficzne w Nauce (2005), no.~37, 73--97.

\bibitem{mitc96}
M.~Mitchell, \emph{An introduction to genetic algorithms}, MIT Press,
  Cambridge, MA, 1996.

\bibitem{moore1990unpredictability}
Cristopher Moore, \emph{Unpredictability and undecidability in dynamical
  systems}, Physical Review Letters \textbf{64} (1990), no.~20, 2354.

\bibitem{moore2011nature}
Cristopher Moore and Stephan Mertens, \emph{The nature of computation}, Oxford
  University Press, 2011.

\bibitem{moschovakis2010kleene}
Yiannis~N Moschovakis, \emph{Kleene's amazing second recursion theorem},
  Bulletin of Symbolic Logic \textbf{16} (2010), no.~2, 189--239.

\bibitem{nielsen1997computable}
Michael~A Nielsen, \emph{Computable functions, quantum measurements, and
  quantum dynamics}, Physical Review Letters \textbf{79} (1997), no.~15, 2915.

\bibitem{nielsen2010quantum}
Michael~A Nielsen and Isaac~L Chuang, \emph{Quantum computation and quantum
  information}, Cambridge university press, 2010.

\bibitem{nielsen2006quantum}
Michael~A Nielsen, Mark~R Dowling, Mile Gu, and Andrew~C Doherty, \emph{Quantum
  computation as geometry}, Science \textbf{311} (2006), no.~5764, 1133--1135.

\bibitem{parberry2013knowledge}
Ian Parberry, \emph{Knowledge, understanding, and computational complexity},
  Optimality in biological and artificial networks?, Routledge, 2013,
  pp.~141--160.

\bibitem{piccinini2011physical}
Gualtiero Piccinini, \emph{The physical church--turing thesis: Modest or
  bold?}, The British Journal for the Philosophy of Science (2011).

\bibitem{piccinini2010computation}
Gualtiero Piccinini and Corey Maley, \emph{Computation in physical systems},
  (2010).

\bibitem{pour1982noncomputability}
Marian~Boykan Pour-El and Ian Richards, \emph{Noncomputability in models of
  physical phenomena}, International Journal of Theoretical Physics \textbf{21}
  (1982), no.~6, 553--555.

\bibitem{raattkainen2005philosophical}
Panu Raattkainen, \emph{On the philosophical relevance of godel's
  incompleteness theorems}, Revue internationale de philosophie \textbf{59}
  (2005), no.~4, 513--534.

\bibitem{savage1998models}
John~E Savage, \emph{Models of computation}, vol. 136, Addison-Wesley Reading,
  MA, 1998.

\bibitem{shiraishi2021undecidability}
Naoto Shiraishi and Keiji Matsumoto, \emph{Undecidability in quantum
  thermalization}, Nature communications \textbf{12} (2021), no.~1, 1--7.

\bibitem{shore2016Turing}
Richard~A Shore, \emph{The turing degrees: an introduction}, Forcing, iterated
  ultrapowers, and Turing degrees, World Scientific, 2016, pp.~39--121.

\bibitem{sipser1996introduction}
Michael Sipser, \emph{Introduction to the theory of computation}, ACM Sigact
  News \textbf{27} (1996), no.~1, 27--29.

\bibitem{sipser2006introduction}
\bysame, \emph{Introduction to the theory of computation}, vol.~2, Thomson
  Course Technology Boston, 2006.

\bibitem{soare2016Turing}
Robert~I Soare, \emph{Turing computability: Theory and applications}, vol. 300,
  Springer, 2016.

\bibitem{tegmark1998theory}
Max Tegmark, \emph{Is ``the theory of everything'' merely the ultimate ensemble
  theory?}, Annals of Physics \textbf{270} (1998), no.~1, 1--51.

\bibitem{tegmark2008mathematical}
\bysame, \emph{The mathematical universe}, Foundations of Physics \textbf{38}
  (2008), no.~2, 101--150.

\bibitem{watson2003zhuangzi}
Burton Watson et~al., \emph{Zhuangzi: Basic writings}, Columbia University
  Press, 2003.

\bibitem{weatherson2003you}
Brian Weatherson, \emph{Are you a sim?}, The Philosophical Quarterly
  \textbf{53} (2003), no.~212, 425--431.

\bibitem{wikipedia_state_transition_system}
Wikipedia, \emph{State transition systems}, December 2023.

\bibitem{wolfram1984cellular}
Stephen Wolfram, \emph{Cellular automata as models of complexity}, Nature
  \textbf{311} (1984), no.~5985, 419--424.

\bibitem{wolpert2024know}
David~H. Wolpert, \emph{What can we know about that which we cannot even
  imagine?}, 2024, In press in ``New Frontiers in Science", Marilena
  Streit-Bianchi and Vittorio Gorini (Ed.).

\bibitem{zure89b}
W.~H. Zurek, \emph{Algorithmic randomness and physical entropy}, Phys. Rev. A
  \textbf{40} (1989), 4731--4751.

\end{thebibliography}

\end{document}